\documentclass[ba,preprint]{imsart}

\RequirePackage{amsthm,amsmath,amsfonts,amssymb}
\RequirePackage[authoryear]{natbib}
\RequirePackage[colorlinks,citecolor=blue,urlcolor=blue,backref=page,backref=page]{hyperref}
\RequirePackage{graphicx}
\usepackage{subfigure}
\usepackage{physics}
\usepackage{algorithm}

\usepackage{appendix}

\usepackage{algpseudocode}
\usepackage[capitalize]{cleveref}

\startlocaldefs
\theoremstyle{plain}

\newtheorem{theorem}{Theorem}[section]

\newtheorem{corollary}{Corollary}[theorem]
\theoremstyle{definition}
\newtheorem{definition}[theorem]{Definition}

\theoremstyle{remark}

\newcommand{\QP}{\texttt{QPMCMC}}

\newcommand{\QPP}{\texttt{QPMCMC2}}

\newcommand{\bt}{\boldsymbol{\theta}}

\endlocaldefs

\begin{document}

\begin{frontmatter}
\title{Quantum Speedups for Multiproposal MCMC}

\begin{aug}
\author[A]{\fnms{Chin-Yi }~\snm{Lin}\ead[label=e1]{cylin1113@gmail.com}},
\author[B]{\fnms{ Kuo-Chin}~\snm{Chen}\ead[label=e2]{Jim.KC.Chen@foxconn.com}},
\author[C]{\fnms{Philippe}~\snm{Lemey}\ead[label=e3]{philippe.lemey@kuleuven.be}},
\author[D]{\fnms{Marc}~\snm{A.~Suchard}\ead[label=e4]{msuchard@ucla.edu}},
\author[E]{\fnms{Andrew}~\snm{J.~Holbrook}\ead[label=e5]{aholbroo@g.ucla.edu}},
\and
\author[B]{\fnms{Min-Hsiu}~\snm{Hsieh}\ead[label=e6]{min-hsiu.hsieh@foxconn.com}}







\address[A]{Department of Physics, National Taiwan University, Taipei, Taiwan\printead[presep={,\ }]{e1}}

\address[B]{Foxconn Research, Taipei, Taiwan\printead[presep={,\ }]{e2}\printead[presep={,\ }]{e6}}

\address[C]{Department of Microbiology, Immunology and Transplantation, Rega Institute, KU Leuven, Leuven, Belgium\printead[presep={,\ }]{e3}}

\address[D]{Departments of Biostatistics, Biomathematics and Human Genetics, UCLA, Los Angeles, USA\printead[presep={,\ }]{e4}}

\address[E]{Department of Biostatistics, UCLA, Los Angeles, USA\printead[presep={,\ }]{e5}}

\end{aug}

\begin{abstract}
Multiproposal Markov chain Monte Carlo (MCMC) algorithms choose from multiple proposals to generate their next chain step in order to sample from challenging target distributions more efficiently. 
However, on classical machines, these algorithms require $\mathcal{O}(P)$ target evaluations for each Markov chain step when choosing from $P$ proposals.
Recent work demonstrates the possibility of quadratic quantum speedups for one such multiproposal MCMC algorithm. 
After generating $P$ proposals, this quantum parallel MCMC (\QP) algorithm requires only $\mathcal{O}(\sqrt{P})$ target evaluations at each step, outperforming its classical counterpart. However, generating $P$ proposals using classical computers still requires $\mathcal{O}(P)$ time complexity, resulting in the overall complexity of \QP\ remaining $\mathcal{O}(P)$. Here, we present a new, faster quantum multiproposal MCMC strategy, \QPP. With a specially designed Tjelmeland distribution that generates proposals close to the input state, \QPP\ requires only $\mathcal{O}(1)$ target evaluations and $\mathcal{O}(\log P)$ qubits when computing over a large number of proposals $P$. 
Unlike its slower predecessor, the \QPP\ Markov kernel (\textcolor{red}{1}) maintains detailed balance exactly and (\textcolor{red}{2}) is fully explicit for a large class of graphical models.  We demonstrate this flexibility by applying \QPP\ to novel Ising-type models built on bacterial evolutionary networks and obtain significant speedups for Bayesian ancestral trait reconstruction for 248 observed salmonella bacteria. 
\end{abstract}



\end{frontmatter}

\section{Introduction}

\label{sec:intro}

In their many forms, multiproposal MCMC methods \citep{Tjelmeland2004UsingAM,frenkel2004speed,delmas2009does,neal2011mcmc,calderhead2014general,luo2019multiple} use multiple proposals to gain advantage over traditional MCMC algorithms \citep{metropolis1953equation,hastings} that only generate a single proposal at each step. After generating a number of proposals, these methods randomly select the next Markov chain state from a set containing all $P$ proposals and the current state with selection probabilities involving the target and proposal density (mass) functions. However, this claimed advantage has one shortcoming: calculation of proposal probabilities typically scales $\mathcal{O}(P^2)$ which outweighs the aforementioned advantages,
ultimately resulting in degraded performance.
Recent efforts \citep{glatt2022parallel,holbrook2023generating} focus on efficient joint proposal structures that lead to computationally efficient $\mathcal{O}(P)$-time proposal selection probabilities. Even after incorporating efficient joint proposals such as the Tjelmeland correction (Section \ref{sec:multipropMCMC}), selection probabilities still require evaluation of the function at each of the $P$ proposals. 
\cite{holbrook2023quantum} uses the Gumbel-max trick to turn the proposal selection task into a discrete optimization procedure amenable to established quantum optimization techniques \citep{durr1996quantum,yoder2014fixed}. On the one hand, the resulting \QP \ algorithm facilitates quadratic speedups, only requiring $\mathcal{O}(\sqrt{P})$ target evaluations. Although these quadratic speedups are significant, they are still not sufficient for \QP\ to provide advantages when increasing the proposal number $P$. On the other hand, these target evaluations take the form of generic oracle calls embedded within successive Grover iterations \citep{grover1996fast}, the circuit depth of which is not clear.  Worse still, the fact that the optimization algorithms of \citet{durr1996quantum,yoder2014fixed} sometimes fail to obtain the optimum means that the \QP \ Markov kernel fails to maintain detailed balance with non-negligible probability. The relationship between the algorithm's stationary distribution (if it exists) and the target distribution is unclear as a result. 

Our \QPP \ algorithm (Section \ref{sec:qp2}) combines multiproposal MCMC with quantum computing but improves upon \QP \ in multiple ways.  First, the \QPP\ circuit depth is $\mathcal{O}(1)$,  i.e., it does not grow with the number of proposals $P$.  Second, the \QPP\ Markov kernel maintains detailed balance exactly, so the algorithm obtains ergodicity and provides asymptotically exact estimators with the usual guarantees \citep{tierney1994markov}. Third, the \QPP\ circuit is fully explicit for a large class of graphical models, making it possible to quantify circuit depth and the $\mathcal{O}(\log P)$ circuit width.  Our algorithm uses the same efficient multiproposal structures as \QP\ to simplify selection probabilities, but this is where similarities cease.  Instead of indirectly choosing the next Markov chain state via quantum optimization, we directly obtain selection probabilities as quantum probability amplitudes that provide weights for superposed proposal states.  Collapsing the quantum state results in easy proposal selection.

Beyond \QP, other quantum-accelerated MCMC algorithms, such as quantum simulated annealing (QSA) \citep{PhysRevLett.101.130504} and the quantum Metropolis solver (QMS) \citep{montanaro2015quantum, campos2023quantum}, have been proposed. While primarily designed for optimization tasks, these algorithms can also perform sampling. Theoretically, they achieve significant speedups during the convergence process by leveraging quantum phase estimation (QPE) \citep{dorner2009optimal} and Szegedy's quantum walk \citep{szegedy2004quantum}. However, these methods execute all iterations within a single quantum circuit before measurement, drawing only one sample from the target distribution. This approach has two notable limitations: (1) the reliance on very deep quantum circuits, which are susceptible to substantial errors, and (2) the inability to retain the samples generated across iterations as classical data.

Although making a direct comparison between \QPP\ and quantum MCMC algorithms like QSA or QMS is challenging, \QPP\ offers distinct advantages. Unlike QSA and QMS, which rely on a single quantum circuit to process all iterations, \QPP\ employs a separate quantum circuit for each iteration. This significantly reduces the circuit depth and, consequently, the error rates, making \QPP\ more practical for implementation on noisy quantum devices in the near term. Furthermore, the iterative structure of \QPP\ enables the storage of all samples classically after each iteration, offering greater flexibility and usability compared to the designs of QSA and QMS, where only a single sample is generated at the end of the process.

We apply \QPP\ to ancestral trait reconstruction on bacterial evolutionary networks, the irregularity of which serves as a naturally arising test of the algorithm's flexibility.  Phylogenetic comparative methods \citep{felsenstein1985phylogenies} investigate the shared evolution of biological traits and their mutual associations within or across species. Recent statistical efforts in comparative phylogenetics emphasize big data scalability and the application of increasingly complex models that condition on---or jointly infer---phylogenetic trees describing shared evolutionary histories between observed taxa \citep{hassler2023data}. For example,  \cite{zhang2021large,zhang2023accelerating} develop a statistical computing framework for learning dependencies between high-dimensional discrete traits and apply their methods to the Bayesian analysis of, e.g., nearly 1,000 H1N1 influenza viruses. Unfortunately, these methods are ill-suited for bacterial ancestral trait reconstruction.   First, their dynamic programming routines for fast likelihood and log-likelihood gradient calculations rely on the tree structure that directly characterizes the shared evolutionary history of the observed specimens, and the phylogenetic tree fails to capture the reticulate evolution that arises from the exchange of genetic material between microbes. Second, the methods of \cite{zhang2021large,zhang2023accelerating} rely on Gaussianity assumptions in order to efficiently integrate over unobserved ancestral traits and obtain a reduced likelihood describing only the traits of observed specimens.

Given these shortcomings, we instead define novel Ising-type models on Neighbor-Net phylogenetic networks \citep{neighborNet} that directly account for bacterial reticulate evolution.  Within these models, exterior nodes represent observed bacteria, internal nodes represent unobserved ancestors, and spins, the discrete binary variables associated with each node, represent biological traits.  Bayesian ancestral trait reconstruction then amounts to sampling interior spins while keeping exterior spins fixed. We apply our \QPP\ to this sampling task for single- and multi-trait Ising models that arise from a Neighbor-Net network characterizing the evolutionary history shared by 248 salmonella bacteria.  Notably, this same microbial collection features prominently in high-impact studies \citep{mather2013distinguishable,cybis2015assessing} of the evolution and development of antibiotic resistances in salmonella bacteria, a matter of pressing societal concern.

\section{Preliminaries}

We present limited introductions to the methods and ideas that are central to our development and exposition of \QPP, including MCMC, multiproposal MCMC, quantum computing and our Ising-type phylogenetic network models.

\subsection{MCMC and Barker's algorithm}\label{sec:MCMC and Barker's algorithm}

\newcommand{\ttheta}{\boldsymbol{\theta}}

Markov Chain Monte Carlo (MCMC) constitutes a class of algorithms that are useful for sampling from probability distributions in situations where direct sampling is otherwise untenable. Key applications of MCMC include inference of high-dimensional model parameters within Bayesian inference \citep{gelman1995bayesian} and simulation of physical many-body systems \citep{metropolis1953equation,DUANE1987216}. In the following, we consider the application of MCMC to discrete-valued models, but the framework applies equally to both discrete and continuous contexts. Letting $\mathcal{A}$ denote some finite or countably-infinite index set, we consider the discrete set $\{\ttheta_{\alpha} \}_{\alpha \in \mathcal{A}}$. 
We identify our target distribution $\pi$ with a probability mass function $\pi(\cdot)$ defined with respect to the counting measure on the power set $2^{\mathcal{A}}$. The probability measure $\pi$ may be, e.g., a posterior distribution in Bayesian inference or a Boltzmann distribution in statistical mechanics. 
However, the probability mass function $\pi(\cdot)$ cannot be accessed in most practical scenarios. Instead, an unnormalized function $\pi^*(\cdot)\propto \pi(\cdot)$ is accessible. 

In this context, Monte Carlo methods generate (pseudo) random samples in order to obtain estimates of expectations $E_{\pi}(f)< \infty$ for arbitrary bounded functions $f$ defined on the set $\{\ttheta_{\alpha} \}_{\alpha \in \mathcal{A}}$.  Whereas classical Monte Carlo techniques such as rejection sampling tend to break down in high dimensions, MCMC effectively generates samples from high-dimensional distributions by constructing a Markov chain with transition kernel $Q(\cdot,\cdot)$ that maintains the target distribution $\pi$ as a stationary distribution, i.e.,
\begin{align}\label{eq:stationary}
    \pi(\alpha) = \sum_{\alpha'} \pi(\alpha') Q(\alpha',\alpha), \quad \forall \alpha \in \mathcal{A} \, .
\end{align}
When designing such Markov kernels $Q$, it is helpful to note that the detailed balance condition
\begin{align}
   \pi(\alpha') Q(\alpha',\alpha) =  \pi(\alpha) Q(\alpha,\alpha') , \quad \forall \alpha, \alpha' \in \mathcal{A}
\end{align}
 guarantees the kernel $Q$'s satisfaction of \eqref{eq:stationary}, while at the same time verifying more easily than \eqref{eq:stationary}.  The Metropolis-Hastings kernel \citep{metropolis1953equation} maintains detailed balance using two steps: first, it generates a random proposal $\ttheta_1 \sim q(\ttheta_0,\ttheta_1)$, where $\ttheta_0:=\ttheta^{(s-1)}$ is the current state of the Markov chain; second, it accepts the proposal with probability $a_{MH}(\ttheta_0,\ttheta_1)$ or remains in the current state for one more iteration. 

In fact, other acceptance probabilities besides $a_{MH}$ also maintain detailed balance when coupled with proposals of the form $q(\ttheta_0,\ttheta_1)$. We are particularly interested in the Barker \citep{barker1965monte} acceptance probability
\begin{equation}\label{eq:barkerAccept}
 a_B := \frac{\pi(\boldsymbol{\theta}_p)q(\boldsymbol{\theta}_p, \boldsymbol{\theta}_{|p-1|})}{\sum_{p'=0}^{1}\pi(\boldsymbol{\theta}_{p'})q(\boldsymbol{\theta}_{p'}, \boldsymbol{\theta}_{|p'-1|})}, \quad p\in \{0,1\}.
\end{equation}
When $q(\cdot,\cdot)$ is symmetric in its two arguments, \eqref{eq:barkerAccept} takes the salient form
\begin{equation}\label{eq:barkerSimple}
 \frac{\pi(\boldsymbol{\theta}_p)}{\sum_{p'=0}^{1}\pi(\boldsymbol{\theta}_{p'})} =:\pi_p, \quad p\in \{0,1\},
\end{equation}
leading to Algorithm \ref{alg:barker}.  The notation of \eqref{eq:barkerAccept} and \eqref{eq:barkerSimple} extends to the multiple proposal case. Here, the development of symmetric joint proposals and simplified acceptances $\pi_p$ is not straightforward, but leads to significant computational efficiencies.
It is worth noting that the simplified acceptances $\pi_p$ can be obtained by replacing $\pi$ with $\pi^{*}$:

\begin{equation}
     \frac{\pi^*(\boldsymbol{\theta}_p)}{\sum_{p'=0}^{1}\pi^*(\boldsymbol{\theta}_{p'})} = \frac{\pi(\boldsymbol{\theta}_p)}{\sum_{p'=0}^{1}\pi(\boldsymbol{\theta}_{p'})} =\pi_p, \quad p\in \{0,1\}.
\end{equation}

\begin{algorithm}[!t]
\caption{MCMC with Barker Acceptances and Symmetric Proposals}
\label{alg:barker}
\begin{algorithmic}[1] 
\Require         
An initial Markov chain state $\boldsymbol{\theta}^{(0)}$; a routine
for evaluating a function $\pi^{*}(\cdot) \propto \pi(\cdot)$, where $\pi(\cdot)$ is our target distribution's probability mass function; a routine for sampling $\ttheta'$ from a proposal distribution $q(\boldsymbol{\theta}, \boldsymbol{\theta}')$ symmetric in $\ttheta$ and $\ttheta'$; a routine for sampling from a discrete distribution $\operatorname{Discrete}(\cdot)$ parameterized by an arbitrary probability vector; the number of samples to generate $S$. 
\For{$s \in \{1,\ldots,S\} $}
\State $\boldsymbol{\theta_0} \leftarrow \boldsymbol{\theta}^{(s-1)}$; $\boldsymbol{\theta}_{1}\sim q(\boldsymbol{\theta}_0,\cdot)$;
\State $\boldsymbol{\pi^{*}}=(\pi^{*}_0,\pi^{*}_1)^T$ where
$\pi^{*}_0 \leftarrow \pi^{*}(\boldsymbol{\theta_0})$ and $\pi^{*}_1 \leftarrow \pi^{*}(\boldsymbol{\theta_1})$;

\State$\hat{p} \sim \operatorname{Discrete}(\boldsymbol{\pi}^{*}/\boldsymbol{\pi}^{*T}\boldsymbol{1})$; $\boldsymbol{\theta^{(s)}} \leftarrow \boldsymbol{\theta_{\hat{p}}}$;
\EndFor 

\State \Return $\boldsymbol{\theta^{(1)}},\ldots,\boldsymbol{\theta^{(S)}}$.
\end{algorithmic} 
\end{algorithm}

\subsection{Multiproposal MCMC and the Tjelmeland correction}\label{sec:multipropMCMC}

\newcommand{\TTheta}{\boldsymbol{\Theta}}

Multiproposal MCMC algorithms use multiple proposals at each iteration to explore target distributions more efficiently.  Recently, \cite{glatt2022parallel} present general measure theoretic foundations for the many different multiproposal MCMC algorithms that already exist. Among many other important contributions, this abstract multiproposal MCMC framework incorporates: both Metropolis-Hastings-type and Barker-type multiproposal MCMC acceptance criteria; and efficient joint proposal structures \citep{Tjelmeland2004UsingAM,holbrook2023generating} called Tjelmeland corrections.  We follow \citet{holbrook2023generating,holbrook2023quantum} and consider a multiproposal MCMC algorithm that combines Barker-type acceptance criteria with the Tjelmeland correction.

Again letting \( \boldsymbol{\theta}_0 :=\ttheta^{(s-1)} \) denote the current state of the Markov chain, one version of multiproposal MCMC proceeds by generating $P$ proposals $(\ttheta_1,\dots,\ttheta_P)=:\TTheta_{-0}$ from some joint distribution with probability mass function $q(\ttheta_0,\TTheta_{-0})$ and randomly selecting the next Markov chain state from among the current and proposed states with probabilities 

\begin{equation} \label{parallel pi_p}
\pi_p := \frac{\pi(\boldsymbol{\theta}_p)q(\boldsymbol{\theta}_p, \boldsymbol{\Theta}_{-p})}{\sum_{p'=0}^{P}\pi(\boldsymbol{\theta}_{p'})q(\boldsymbol{\theta}_{p'}, \boldsymbol{\Theta}_{-p'})}= \frac{\pi^*(\boldsymbol{\theta}_p)q(\boldsymbol{\theta}_p, \boldsymbol{\Theta}_{-p})}{\sum_{p'=0}^{P}\pi^*(\boldsymbol{\theta}_{p'})q(\boldsymbol{\theta}_{p'}, \boldsymbol{\Theta}_{-p'})}, \quad p\in \{0,\dots,P \},
\end{equation}
where $\TTheta_{-p}$ is the $P$-columned matrix that results when one extracts the vector $\ttheta_p$ from the matrix $(\ttheta_0,\ttheta_1,\dots,\ttheta_P)$.  Given the burdensome $\mathcal{O}(P^2)$ floating-point operations required to evaluate all $P+1$ joint mass functions $q(\ttheta_p,\TTheta_{-p})$, \cite{holbrook2023generating} recommends using joint proposal strategies that enforce the higher-order symmetry relation
\begin{align}\label{eq:highSym}
    q(\ttheta_0,\TTheta_{-0}) = q(\ttheta_1,\TTheta_{-1}) = \cdots = q(\ttheta_P,\TTheta_{-P}) 
\end{align}
and lead to simplified acceptance probabilities
\begin{equation}\label{eq:barkerMulti}
\pi_p = \frac{\pi^*(\boldsymbol{\theta}_p)}{\sum_{p'=0}^{P}\pi^*(\boldsymbol{\theta}_{p'})} , \quad p\in \{0,1,\dots,P\}.
\end{equation}
To this end, \cite{holbrook2023generating} shows that an elegant joint proposal structure put forth by \cite{Tjelmeland2004UsingAM} leads to \eqref{eq:highSym}.  This Tjelmeland correction uses a symmetric probability distribution with mass function satisfying $\bar{q}(\ttheta,\ttheta')=\bar{q}(\ttheta',\ttheta)$ to first generate a random offset $\bar{\ttheta} \sim \bar{q}(\ttheta_0,\cdot)$ and then generate $P$ proposals $\ttheta_1,\dots,\ttheta_P\stackrel{iid}{\sim}\bar{q}(\bar{\ttheta},\cdot)$.
Because
\begin{align*}
    q(\ttheta_0,\TTheta_{-0}) = \sum_{\bar{\ttheta}} \bar{q}(\ttheta_0,\bar{\ttheta}) \prod_{p'\neq 0} \bar{q}(\bar{\ttheta},\ttheta_{p'}) =
    \sum_{\bar{\ttheta}} \bar{q}(\ttheta_p,\bar{\ttheta}) \prod_{p'\neq p} \bar{q}(\bar{\ttheta},\ttheta_{p'}) = q(\ttheta_p,\TTheta_{-p}) \, ,
\end{align*}
this joint proposal strategy satisfies \eqref{eq:highSym} and leads to the simple multiproposal MCMC routine shown in Algorithm \ref{alg:multiProp_iter}.  In the following, we refer to $\bar{q}(\cdot,\cdot)$ as a Tjelmeland kernel and $\bar{q}(\ttheta,\cdot)$ as a Tjelmeland distribution. According to Theorem 2.11 and Corollary 2.12 in \cite{10.1093/imatrm/tnae004}, \cref{alg:multiProp} is guaranteed to maintain detail balance and leaves the target distribution $\pi^*(\cdot)$ invariant.




\begin{algorithm}[!t]
\caption{Multiproposal MCMC with Barker Acceptances and the Tjelmeland Correction}
\label{alg:multiProp}
\begin{algorithmic}[1] 
\Require         
An initial Markov chain state $\boldsymbol{\theta}^{(0)}$;  a routine
for evaluating a function $\pi^{*}(\cdot) \propto \pi(\cdot)$, where $\pi(\cdot)$ is our target distribution's probability mass function; a routine for sampling $\ttheta'$ from a Tjelmeland distribution $\bar{q}(\boldsymbol{\theta}, \boldsymbol{\theta}')$ symmetric in $\ttheta$ and $\ttheta'$; a routine for sampling from a discrete distribution $\operatorname{Discrete}(\cdot)$ parameterized by an arbitrary probability vector; the number of samples to generate $S$; the number of proposals $P$. 
\For{$s \in \{1,\ldots,S \} $}
\State $\boldsymbol{\theta_0} \leftarrow \boldsymbol{\theta}^{(s-1)}$;
\State $\boldsymbol{\theta^{(s)}} \leftarrow $ Multiproposal MCMC iteration (\cref{alg:multiProp_iter})
\EndFor 

\State \Return $\boldsymbol{\theta^{(1)}},\ldots,\boldsymbol{\theta^{(S)}}$.
\end{algorithmic} 
\end{algorithm}

\begin{algorithm}[!t]
\caption{ Multiproposal MCMC iteration
with Barker Acceptances and the Tjelmeland Correction}
\label{alg:multiProp_iter}
\begin{algorithmic}[1] 
\Require         
An input Markov chain state $\boldsymbol{\theta}_0$; a routine
for evaluating a function $\pi^{*}(\cdot) \propto \pi(\cdot)$, where $\pi(\cdot)$ is our target distribution's probability mass function; a routine for sampling $\ttheta'$ from a Tjelmeland distribution $\bar{q}(\boldsymbol{\theta}, \boldsymbol{\theta}')$ symmetric in $\ttheta$ and $\ttheta'$; a routine for sampling from a discrete distribution $\operatorname{Discrete}(\cdot)$ parameterized by an arbitrary probability vector; the number of proposals $P$. 

\State $\bar{\boldsymbol{\theta}}\sim \bar{q}(\boldsymbol{\theta}_0,\cdot)$; $\ttheta_1,\dots,\ttheta_P\stackrel{iid}{\sim} \bar{q}(\bar{\ttheta},\cdot)$;
\State $\boldsymbol{\pi^*}=(\pi^*_0,\pi^*_1, \dots,\pi^*_P)^T$ where
$\pi_p^{*} \leftarrow \pi^*(\boldsymbol{\theta_p}), \, p \in \{0,1,\dots, P\}$;

\State$\hat{p} \sim \operatorname{Discrete}(\boldsymbol{\pi^*}/\boldsymbol{\pi^*}^T\boldsymbol{1})$; 

\State \Return $\boldsymbol{\theta_{\hat{p}}}$.
\end{algorithmic} 
\end{algorithm}

\subsection{An introduction to quantum computing}

In this section, we provide an introduction to quantum computing, consisting of three parts. First, we introduce quantum bits (qubits), which are analogous to classical bits in storing information. Second, we present the unitary operator, which acts as a quantum logic gate to manipulate qubits. Finally, we explain the rules of measurement in quantum computing.

\subsubsection{Qubits and quantum states}

In quantum mechanics, the state of a physical system is represented by a vector in a Hilbert space, which is a complex vector space with an inner product. Bra-ket notation is used to denote these vectors, where \(\ket{v}\), known as a ket, identifies the vector in this complex space. The corresponding bra, \(\bra{v}\), represents the complex conjugate transpose of \(\ket{v}\). The inner product of two vectors, \(\ket{v}\) and \(\ket{w}\), is written as \(\braket{v}{w}\).

In quantum information, the computational basis vectors are represented by \(\ket{0}\) and \(\ket{1}\). These vectors are assumed to be normalized and orthogonal to each other. The notation \(\ket{0}\) and \(\ket{1}\) is analogous to classical bits, which take the values of $0$ or $1$. In quantum information theory, these states exist within a two-dimensional Hilbert space, representing the possible states of a quantum bit, or qubit. A qubit exists in a superposition of these states, such as \[\ket{\psi}=\psi_0\ket{0}+\psi_1\ket{1},\] where $\psi_0$ and $\psi_1$ are complex numbers satisfying |$\psi_0|^2 + |\psi_1|^2 = 1$. This superposition describes a valid quantum state within the Hilbert space.

When dealing with a system comprising more than one qubit, the state of the entire system is in the tensor product space of the individual qubit states. For an $n$-qubit system, a basis state can generally be expressed as
$$\ket{q_1}\otimes\ket{q_2}\otimes \dots \otimes \ket{q_n}=\ket{q_1}\ket{q_2}\dots \ket{q_n}=
\ket{q_1, q_2, \dots, q_n},$$
where each $q_i \in \{0, 1\}$, and for simplicity, the tensor product symbol $\otimes$ is often omitted. An $n$-qubit system has $2^n$ possible basis states. The basis state can also be labeled by an integer $\sum_{i=0}^{n-1}2^{i}q_i$.
An $n$-qubit quantum state can then be expressed as a linear combination of these basis states \(\ket{k}\) for \(k \in \{0, \dots, 2^n-1\}\), as shown below:
\begin{equation} \label{eq:quantum_state}
    \ket{\psi}=\sum_{k=0}^{2^n-1}\psi_k\ket{k},
\end{equation}
where $\psi_k$ are complex coefficients $(\psi_k \in \mathbb{C})$ that satisfy the normalization condition $\sum_{k=0}^{2^n-1}|\psi_k|^2=1$.

We refer to a collection of qubits as a quantum register, and denote the register label as a subscript on the quantum state. For instance, if we have $n + m$ qubits, the system of $n$ qubits can be labeled as $\mathcal{X}$ and the system of $m$ qubits as $\mathcal{Y}$. The basis state of this $n + m$ qubit system $\ket{q_0, \dots,q_{n+m-1}}$ can then be expressed as:
$$
\ket{x_0, \dots,x_{n-1}}_{\mathcal{X}}\ket{y_{0}, \dots, y_{m-1}}_{\mathcal{Y}}=\ket{x}_{\mathcal{X}}\ket{y}_{\mathcal{Y}},
$$
where $x_{j}=q_{j}$ and $y_{k}=q_{n+k}$, with \(x \in \{0, \dots, 2^n - 1\}\) and \(y \in \{0, \dots, 2^m - 1\}\).

\subsubsection{Quantum operations}\label{sec:Quantum operations}

In the domain of quantum computation, qubits are manipulated through the application of unitary operators to quantum states. A unitary operator $O$ is a linear operator on a Hilbert space that satisfies  $OO^{\dagger}=O^{\dagger}O=I$, where  $O^{\dagger}$ represents the complex conjugate transpose of  $O$, and $I$  is the identity operator in this Hilbert space.

A unitary operator $O$  acts upon an  $n$ qubit quantum superposition state \emph{in parallel}, following the equation:
$$
O\ket{\psi}= O\sum_{k=0}^{2^n-1}\psi_k\ket{k} = \sum_{k=0}^{2^n-1} \psi_k O\ket{k}.
$$


To implement a unitary operator, we must synthesize it using a universal set of quantum gates, which can be categorized into three main classes:

\begin{itemize}
\item 
Phase Shift Gate: This gate maps the basis states as \( \ket{0} \mapsto \ket{0} \) and \( \ket{1} \mapsto e^{i\varphi} \ket{1} \), and it can be represented as: $P(\phi)=\begin{bmatrix}
1& 0 \\
0 & \exp(i\phi)
\end{bmatrix}$ when we denote $\ket{0}$ as $\begin{bmatrix}
1 \\
0 \end{bmatrix}$, and $\ket{1}$ as $\begin{bmatrix}
0 \\
1 \end{bmatrix}$.
\item 
Rotation Gates: These gates apply rotations around specific axes. They can be represented as:
\begin{align*}
R_x(\theta)&=\begin{bmatrix}
\cos(\theta/2) & -i \sin(\theta/2) \\
-i \sin(\theta/2) & \cos(\theta/2)
\end{bmatrix}, 
R_y(\theta)=
\begin{bmatrix}
\cos(\theta/2) & -\sin(\theta/2) \\
\sin(\theta/2) & \cos(\theta/2)
\end{bmatrix}, \text{and}
\\
R_z(\theta)&=
\begin{bmatrix}
\exp(-i\theta/2) & 0 \\
0 & \exp(i\theta/2)
\end{bmatrix}.
\end{align*}
\item 
Controlled-NOT Gate (CNOT): It operates on two qubits: a control qubit and a target qubit. The CNOT gate flips the state of the target qubit if the control qubit is in the \(\ket{1}\) state; otherwise, the target qubit remains unchanged.
For example, in a two-qubit system \(\ket{q_1, q_2}\), where $q_1$ is the control qubit and $q_2$ is the target qubit, the CNOT gate performs the following operations:

\[
\begin{array}{c|c|c}
q_1 & q_2 & \text{CNOT} \\
\hline
0 & 0 & 0 \\
0 & 1 & 1 \\
1 & 0 & 1 \\
1 & 1 & 0 \\
\end{array}
\]

\end{itemize}

Each general quantum machine can utilize these quantum gates, constrained to specific choices of \(\theta\) and \(\phi\). We denote \(\mathbb{T}(O)\) as the number of gates required to synthesize a unitary operation \(O\). In the context of a quantum circuit, the circuit depth is defined as the length of the longest path from the input (or from a state preparation) to the output, measured in terms of gate applications.

Two specific unitary operations are of interest in this study. First, we introduce a unitary operation that implements any classical function within a quantum computer, as outlined by \citep{nielsen2010quantum}.

Given a well-defined function \(f: X \rightarrow Y\) for two finite sets \(X\) and \(Y\) with their elements encoded in quantum registers $ \mathcal{X}$ and $\mathcal{Y}$, respectively, a unitary operator \(O_f\) can be constructed.
This operator acts on three registers \(\mathcal{X}\), \(\mathcal{Y}\), and \(\mathcal{W}\) as follows:
$$
O_f\ket{x}_{\mathcal{X}}\ket{0}_{\mathcal{Y}} = \ket{x}_{\mathcal{X}}\ket{f(x)}_{\mathcal{Y}},
$$
where \(x \in X\), \(f(x) \in Y\).
The working register $\mathcal{W}$ assists in computing $f(x)$, though for simplicity, it can be ignored here. The number of gates required to implement $O_f$, denoted as $\mathbb{T}(O_f)$, is asymptotically no greater than the classical time complexity for computing $f(x)$.


Second, we introduce a controlled rotation gate operating on three registers: \(\mathcal{X}\), \(\mathcal{Y}\), and \(\mathcal{C}\):
\begin{itemize}
    \item \(\mathcal{X}\) represents a number \(0 \leq x \leq 1\) to a specified precision.
    \item \(\mathcal{Y}\) contains one qubit, initially set to \(\ket{0}\).
    \item \(\mathcal{C}\) represents a finite set \(C\).
\end{itemize}

The controlled rotation gate maps \(\ket{0}_{\mathcal{Y}}\) to \(\sqrt{1-x}\ket{0}_{\mathcal{Y}} + \sqrt{x}\ket{1}_{\mathcal{Y}}\), conditioned on the state \(\ket{c}_{\mathcal{C}}\) for some $c \in C$. In other words, the gate rotates the qubit in $\mathcal{Y}$ only when the state in the \(\mathcal{C}\) register is \(\ket{c}\).

\subsubsection{Measurement in quantum computing}

In quantum computation, measurement is the process of extracting classical information from a quantum system, causing its quantum state to collapse into one of the possible classical states. This is a critical step in quantum algorithms because it determines the final output after a series of quantum operations.

Before measurement, qubits exist in a superposition of states. For example, a qubit can be in a superposition of \( \ket{0} \) and \( \ket{1} \), and its state can be expressed as \( \ket{\psi} = \psi_0 \ket{0} + \psi_1 \ket{1} \), where $\psi_0, \psi_1 \in \mathbb{C}$  are complex numbers such that $|\psi_0|^2 + |\psi_1|^2 = 1 $.

Upon measurement, the superposition collapses to a single, definite state either \( \ket{0} \) or \( \ket{1} \)    with probabilities determined by the square of the amplitudes of the corresponding quantum states. Specifically, for the superposition state \( \psi_0 \ket{0} + \psi_1 \ket{1} \), the measurement outcome will be \( \ket{0} \) with probability  $|\psi_0|^2$, or \( \ket{1} \) with probability  $|\psi_1|^2 $. After measurement, the qubit’s state collapses to the observed result. If the outcome is $0 $, the qubit becomes \( \ket{0} \), and similarly, if the outcome is $ 1 $, the qubit becomes \( \ket{1} \).

Thus, measurement not only provides the classical result of the quantum computation but also irreversibly alters the qubit’s state by collapsing the superposition.


\section{Main Result: Fast Quantum Parallel MCMC}\label{sec:main}

In this section, we introduce our main result for \QPP \ (\cref{alg:qpmcmc2}), which presents a novel quantum sampling algorithm that surpasses the classical multiproposal MCMC (\cref{alg:multiProp}). 
Subsequently, we delineate the technical contribution, and illustrate its significance in addressing the bottleneck identified in \cref{alg:multiProp}. Finally, a comprehensive elucidation of our algorithm will be presented in detail.

\QPP\ (\cref{alg:qpmcmc2}) is a quantum sampling algorithm that serves as the quantum counterpart to the classical multiproposal MCMC (\cref{alg:multiProp}), offering improved time complexity under specific conditions. This acceleration is achieved by substituting the classical iteration step (\cref{alg:multiProp_iter}) with its quantum-enhanced version (\cref{alg:qpmcmc2_iter}).

\begin{theorem}\label{theorem:main}

\cref{alg:qpmcmc2_iter} is a quantum algorithm that is equivalent to \cref{alg:multiProp_iter}, with the additional input $\mathcal{L}$ that is a constant larger than $\max_{\bt \sim \bar{q}(\bt',\cdot);\bt'\in\mathcal{A} } \left[\frac{\pi(\bt)}{\pi(\bt')}\right]$. This quantum algorithm has a success probability given by:
$$R = \sum_{p \in \{0, \dots, P\}} \frac{\pi_{\bar{\bt}}^*(\boldsymbol{\theta}_p)}{P + 1},$$
where $\bar{\bt} \sim \bar{q}(\bt_0,\cdot)$ is the intermediate state generated from input state $\bt_0$, and  $\pi_{\bar{\bt}}^*:=\frac{\pi}{\pi(\bar{\bt})\mathcal{L}}$ is an unnormalized probability mass function corresponding to $\pi$. 
The time complexity of this quantum algorithm is expressed as:
\begin{align}\label{eq:compl}
2\mathbb{T}(O_{\bar{q}})+\mathbb{T}(O_{\pi_{\bar{\bt}}^*})+\mathcal{O}(1),
\end{align}
where $O_{\bar{q}}$ and $O_{\pi_{\bar{\bt}}^*}$ represent the quantum operations corresponding to $\bar{q}(\bt,\bt')$ and $\pi_{\bar{\bt}}^*(\cdot)$, respectively\footnote{See \cref{sec:Quantum operations} for a detailed description.}.
Furthermore, the success probability $R$ can be lower bounded by a quantity that only depends on $\bar{q}$ and $\pi$ as follow:
\begin{align}\label{eq:LB_R}
R \geq \frac{\mathcal{M}}{\mathcal{L}}
,
\end{align}
where $\mathcal{M}=\min_{\bt \sim \bar{q}(\bt',\cdot);\bt'\in\mathcal{A}}\left[\frac{\pi(\bt)}{\pi(\bt')}\right]$.

\end{theorem}








\cref{alg:qpmcmc2_iter} is the sub-algorithm used in each iteration of \cref{alg:qpmcmc2}, representing a quantum accelerated version of \cref{alg:multiProp_iter}, which is the sub-algorithm of a single iteration of the classical multiproposal MCMC (\cref{alg:multiProp}). We note that \cref{alg:multiProp_iter} requires evaluating $\pi^{*}$ and $\bar{q}$ a total of $P$ times, leading to $\mathcal{O}(P)$ implementations of these functions. 
In contrast, \cref{alg:qpmcmc2_iter} requires only $\mathcal{O}(1)$ implementations, making it independent of the choice of $P$.
Although \cref{alg:qpmcmc2_iter} may fail with a probability $1-R$, it is lower bounded in \cref{eq:LB_R} which is independent of $P$. 
With this quantum-enhanced MCMC iteration (\cref{alg:qpmcmc2_iter}), the time complexity of \QPP\ (\cref{alg:qpmcmc2}) remains independent of $P$. 

Notice that in \cref{alg:qpmcmc2_iter}, the lower bound of the success rate $R$ given in \cref{eq:LB_R} can generally be small while remaining independent of $P$. Here, we demonstrate that for a target distribution $\pi(\cdot)$, if $\log{\pi(\cdot)}$ satisfies the Lipschitz continuity property (\cref{lemma: lipschiz}), it is possible to calculate higher lower bounds for $R$ for certain specifically designed Tjelmeland distributions $\bar{q}(\cdot,\cdot)$.

\begin{definition} \label{lemma: lipschiz}
    Let  $K \in \mathbb{R}^+$  be a positive constant, and let $(\mathcal{X}, d_\mathcal{X})$ and $(\mathcal{Y}, d_\mathcal{Y})$ be two metric spaces, where  $d_\mathcal{X}$ and $d_\mathcal{Y}$ are the metrics on the sets $\mathcal{X}$ and  $\mathcal{Y}$, respectively.
    A function $ f: \mathcal{X} \to \mathcal{Y}$  is said to be  $K$-Lipschitz continuous if:
    $$
    d_\mathcal{Y}(f(x_1), f(x_2)) \leq K \, d_\mathcal{X}(x_1, x_2), \quad \forall x_1, x_2 \in \mathcal{X}.
    $$
\end{definition}

For a target distribution $\pi: \mathcal{A}\rightarrow [0,1]$ such that $\log {\pi(\bt)}$ is $K$-Lipschitz, 
the following property holds: 
\begin{equation} \label{eq:pi_lipschitz}
    \left|\log\left(\frac{\pi(\bt_1)}{\pi(\bt_2)}\right)\right| \leq K d_\mathcal{A}(\bt_1,\bt_2),
\end{equation}
where $\bt_1, \bt_2 \in \mathcal{A}$. 

To optimize the lower bound in \cref{eq:LB_R}, we design the Tjelmeland distribution $\bar{q}(\cdot,\cdot)$ such that it only assigns non-zero probabilities to pairs of states $\bt_1, \bt_2$ that are sufficiently close to each other, with a given threshold $\mathcal{D} \in \mathbb{R}^+$:
\begin{equation} \label{eq:q_design}
   \bar{q}(\bt_1,\bt_2) = 0, \quad \forall d_\mathcal{A}(\bt_1,\bt_2) > \mathcal{D}.
\end{equation}

By combining \cref{eq:pi_lipschitz} and \cref{eq:q_design}, for all $\bt' \in \mathcal{A}$ and $\bt \sim \bar{q}(\bt',\cdot)$, the following inequality holds:
\begin{equation}\label{eq:pi_inequality}
    e^{-KD} \leq \frac{\pi(\bt)}{\pi(\bt')} \leq e^{KD}.
\end{equation}

Thus, by setting $\mathcal{L} = e^{KD}$ and substituting into \cref{eq:LB_R}, we derive:
\[
R \geq 
\dfrac{\min_{\bt \sim \bar{q}(\bt',\cdot);\bt'\in\mathcal{A}} \left[\frac{\pi(\bt)}{\pi(\bt')}\right]}{\mathcal{L}} \geq e^{-2KD}.
\]
From this analysis, we observe that reducing $\mathcal{D}$ leads to a higher lower bound for $R$, thereby guaranteeing a higher success rate for \cref{alg:qpmcmc2_iter}. 

Unlike \cref{alg:multiProp_iter}, \cref{alg:qpmcmc2_iter} requires an additional input 
$$\mathcal{L} \geq \max_{\bt \sim \bar{q}(\bt',\cdot);\bt'\in\mathcal{A}} \left[\frac{\pi(\bt)}{\pi(\bt')}\right].$$ From the above analysis, we have shown that when $\log(\pi(\cdot))$ satisfies \cref{eq:pi_lipschitz}, a better choice of $\mathcal{L} = e^{KD}$ can be achieved. Additionally, in \cref{sec:caseStudy}, we provide an example of how to design $\bar{q}$ to meet the requirement in \cref{eq:q_design} in the case of Ising model sampling. However, it is important to note that there is no guarantee to find such a constant $\mathcal{L}$ for arbitrary distributions $\pi(\cdot)$ and the Tjelmeland distributions $\bar{q}(\cdot, \cdot)$. \cref{alg:qpmcmc2} and \cref{alg:qpmcmc2_iter} is only feasible when $\mathcal{L}$ is provided.

Still, $R$ could be very small in certain cases, leading to an enormous rerun of \cref{alg:qpmcmc2_iter} as the scaling of $\mathcal{O}(1/R)$. In fact, the success rate $R$ in \cref{theorem:main} can be enhanced to surpass $\frac{1}{2}$ with $\mathcal{O}(1/\sqrt{R})$ calls of $O_{\bar{q}}$ and $O_{\pi_{\bar{\bt}}^*}$, using the quantum amplitude amplification algorithm presented by \cite{Brassard:2000xvp}. This improvement leads to the following corollary:

\begin{corollary}\label{cor:AA}
By applying the quantum amplitude amplification algorithm from \cite{Brassard:2000xvp} to \cref{alg:qpmcmc2_iter}, the time complexity  for implementing one iteration in \QPP\ (line 2-5, \cref{alg:qpmcmc2}) is expressed as:
\begin{align}
(2\mathbb{T}(O_{\bar{q}})+\mathbb{T}(O_{\pi_{\bar{\bt}}^*})+\mathcal{O}(1))\cdot\mathcal{O}\left(\frac{1}{\sqrt{R}}\right).
\end{align}
\end{corollary}
The quantum amplitude amplification algorithm introduces a quadratic speedup on the scaling of the success rate $R$, which is a suitable solution to  small $R$. Detailed description of the quantum amplitude amplification algorithm is provided in \cref{sec:AA}.

\subsection{Significance}
The significance of \QPP\ (\cref{alg:qpmcmc2}) is twofold. First, it demonstrates an improvement in time complexity over the classical multiproposal MCMC (\cref{alg:multiProp}) and the earlier quantum parallel MCMC algorithm \QP\ proposed in \citep{holbrook2023quantum}. 
Therefore, \QPP\ can be used to enhance the convergence rate by increasing the number of proposals $P$ without requiring $P$ evaluating $\pi^{*}$ and $\bar{q}$.
Second, it enhances sampling efficiency, particularly by increasing the effective sample size (ESS) \citep{gelman1995bayesian}, leading to more reliable estimates with fewer samples.



\subsubsection{Improvement in time complexity}

Previous studies have demonstrated the advantages of multiproposal MCMC algorithms over, e.g., the Metropolis-Hastings algorithm. Specifically, an increase in the number of proposals $P$ in \cref{alg:multiProp_iter}, which is an iteration of \cref{alg:multiProp}, leads to expedited convergence in the sampling process.

However, this accelerated convergence speed comes with a certain drawback. Typically, the bottleneck in each iteration of the Markov chain, as outlined in \cref{alg:multiProp}, lies in the computation of $\pi^*(\cdot)$. The heightened number of $\pi^*(\cdot)$ computations required by this multiproposal MCMC algorithm demands significant computational resources when augmenting the proposal number $P$ per iteration. In \cref{alg:multiProp}, achieving a single Markov chain iteration through classical computation necessitates a time complexity of $\mathcal{O}(P)$.

Conversely, a quantum circuit can execute parallel calculations across different proposals concurrently. This encompasses the generation of proposal sets, evaluation of $\pi_{\bar{\bt}}^*(\bt_p)$ for each $\bt_{p}$ among this proposal sets, encoding them as a superposition state with $P+1$ components, and the selection of samples among them. This quantum approach holds promise in mitigating the bottleneck of the multiproposal MCMC algorithm, thereby expediting the convergence process. By concurrently processing multiple proposals, this quantum multiproposal MCMC algorithm becomes more competitive in comparison to traditional algorithms that relie on a single proposal.




This work is not the first to utilize quantum circuits in an endeavor to expedite \cref{alg:multiProp}. In a prior investigation \citep{holbrook2023quantum}, the employment of the Grover search approach and the Gumbel-Max trick aimed to devise a quantum algorithm (\QP) for substituting lines 3-4 in~\cref{alg:multiProp}, thereby enhancing the time complexity of these steps from $\mathcal{O}(P)$ to $\mathcal{O}(\sqrt{P})$.
It is noteworthy that, in that study, the acceleration did not extend to the process of generating $P$ proposal sets (line 2, \cref{alg:multiProp}), maintaining the overall complexity for a \QP\ iteration at $\mathcal{O}(P)$.

Upon comparing this work to the previously mentioned study, it becomes evident that our approach signifies a notable advancement over them. When proposal count $P$ is large enough, with a designed Tjelmeland distribution $\bar{q}(\cdot,\cdot)$ that generates proposals closed enough to the input state, we achieve an exponential speedup in terms of the $P$, when contrasted with \cref{alg:multiProp}.

\subsubsection{Improvement in effective sample size}

With \cref{theorem:main}, we can improve another indicator of the sampling efficiency, which is the effective sample size (ESS) \citep{gelman1995bayesian}:
\begin{align*}
    \mbox{ESS} := \frac{S}{\sum_{s=-\infty}^{\infty} \rho(s)} \, ,
\end{align*}
where $S$ is the number of MCMC samples, and $\rho(s)$ is the autocorrelation of a univariate time series at lag $s$.  An effective sampler generally exhibits lower autocorrelation for key model summary statistics, resulting in a larger ESS. 
The ESS provides a measure of how many independent samples the correlated chain is equivalent to: it gives you an idea of the true amount of information your sample contains, taking into account the correlation between sample points. With the time complexity reduction in our quantum algorithm, we are able to achieve a larger effective sample size per oracle, making a more efficient MCMC sampling algorithm.

\subsection{Improved quantum parallel MCMC and its time complexity}\label{sec:qp2}

\begin{algorithm}[!t]
\caption{Quantum accelerated multiproposal MCMC (\QPP)}
\label{alg:qpmcmc2}
\begin{algorithmic}[1] 
\Require         
An initial Markov chain state $\bt^{(0)}$; 
an oracle $O_{\bar{q}}$ for sampling $\ttheta'$ from a Tjelmeland distribution $\bar{q}(\bt, \bt')$ symmetric in $\ttheta$ and $\ttheta'$;
a constant $\mathcal{L} \ge \max_{\bt \sim \bar{q}(\bt',\cdot ) ;\bt'\in\mathcal{A}}[\frac{\pi(\bt)}{\pi(\bt')}]$; a control rotation operator $CR(\cdot)$;
the number of samples to generate $S$; 
the number of proposals $P$.
\For{$s \in \{1,\ldots,S \} $}
\State $\boldsymbol{\theta_0} \leftarrow \boldsymbol{\theta}^{(s-1)}$; $\ttheta^{(s)} \leftarrow \textbf{Stop}$

\While {$\ttheta^{(s)} ==  \textbf{Stop}$}
\State $\ttheta^{(s)} \leftarrow $ Quantum accelerated multiproposal MCMC iteration (\cref{alg:qpmcmc2_iter})
\EndWhile
\EndFor 

\State \Return $\ttheta^{(1)},\ldots,\ttheta^{(S)}$.
\end{algorithmic} 
\end{algorithm}

\begin{algorithm}[!t]
\caption{Quantum accelerated multiproposal MCMC iteration (An iteration of \QPP)}
\label{alg:qpmcmc2_iter}
\begin{algorithmic}[1] 
\Require         
An input Markov chain state $\bt_{0}$;
an oracle $O_{\bar{q}}$ for sampling $\ttheta'$ from a Tjelmeland distribution $\bar{q}(\bt, \bt')$ symmetric in $\ttheta$ and $\ttheta'$; 
an oracle $O_{\pi_{\bt'}^*}$ for evaluating an unnormalized probability mass function  $\pi_{\ttheta'}^*(\cdot) = \frac{\pi(\cdot)}{\pi(\ttheta')\mathcal{L}} \propto \pi(\cdot)$ with the given $\mathcal{L}\ge\max_{\bt \sim \bar{q}(\bt',\cdot ) ;\bt'\in\mathcal{A}}[\frac{\pi(\bt)}{\pi(\bt')}]$;
a control rotation operator $CR(\cdot)$; 
the number of proposals $P$.
\State Prepare a quantum state $\ket{\psi_{0}}=\ket{0}_{\mathcal{P}} \ket{0}_{\mathcal{H}_0}  \ket{0}_{\mathcal{H}_1}  \ket{0}_{\mathcal{H}_2} \ket{0}_{\Pi}\ket{0}_{S}$
\State Encode $\bt_0$ in $\mathcal{H}_0$.
\State Apply $O_{\bar{q}}$, which takes query from $\mathcal{H}_0$ and responses the intermediate state $\bar{\ttheta}$ in $\mathcal{H}_1$
\State Make a uniform superposition state in $\mathcal{P}$
\State Apply $O_{\bar{q}}$, which takes a query from $\mathcal{H}_1$, and responses in $\mathcal{H}_2$ on each state
\State Apply $O_{\pi_{\bar{\bt}}^*}$ with , which takes a query from $\mathcal{H}_2$, and responses in $\Pi$ on each state
\State Apply a control rotation gate $CR$ (controlled by each $\ket{p}_\mathcal{P}$), which takes a query from $\Pi$ and maps $\ket{0}_{\mathcal{S}}$ to $\sqrt{1-\pi_{\bar{\ttheta}}^*(\bt_{p})}\ket{0}_{S}+\sqrt{\pi_{\bar{\ttheta}}^*(\bt_{p})}\ket{1}_{S}$
\State Make a measurement;
\If{$\mathcal{S}$ register is $0$}:
\State \Return \textbf{Stop}
\Else
\State $\bt^{(s)} \leftarrow $ the data in $\mathcal{H}_2$
\State \Return $\bt^{(s)}$
\EndIf 

\end{algorithmic} 
\end{algorithm}

In this subsection, we first provide a detailed description of the quantum-accelerated multiproposal MCMC iteration (\cref{alg:qpmcmc2_iter}) used in \QPP\ (\cref{alg:qpmcmc2}). We then establish its correctness and analyze its time complexity. Similar to \cref{alg:multiProp_iter}, \cref{alg:qpmcmc2_iter} takes the following inputs: an initial Markov chain state $\bt_0$, the number of proposals $P$, and the quantum oracles $O_{\bar{q}}$ and $O_{\pi_{\bar{\bt}}^*}$, which correspond to \( \bar{q}(\bt, \bt') \) and $\pi_{\bar{\bt}}^*$ in \cref{alg:multiProp_iter}, respectively. Additionally, \cref{alg:qpmcmc2_iter} requires an extra input $\mathcal{L}\ge\max_{\bt \sim \bar{q}(\bt',\cdot ) ;\bt'\in\mathcal{A}}[\frac{\pi(\bt)}{\pi(\bt')}]$. Finally, it requires a controlled rotation operation $CR$. These quantum operations are introduced in Section 2.3.

The quantum algorithm begins by initializing several quantum registers according to the following scheme:
\begin{itemize}
    \item The first register, denoted as $\mathcal{P}$, is encoded with the labels of proposals $\{0,\cdots, P\}$ as specified in \cref{alg:multiProp}.
    \item The second register, labeled $\mathcal{H}_0$, is encoded with the input state $\ttheta_0$.
    \item The third register, termed $\mathcal{H}_1$, is encoded with the random offset $\bar{\bt}$ as described in \cref{alg:multiProp}.
    \item The fourth register, denoted as $\mathcal{H}_2$, is encoded with the proposals $\bt_{p} \stackrel{i.i.d.}{\sim} \bar{q}(\bt,\bt')$ for each label of proposal $p\in \{0,\cdots,P\}$.
    \item 
    The fifth register, denoted as $\Pi$, is encoded with the evaluated value from the target distribution $\pi(\cdot)$ for each label of proposal $p$.
    \item The last register, designated as $\mathcal{\mathcal{S}}$, is a register indicating whether the implementation of the Markov chain is successful or not.
\end{itemize}

 The quantum algorithm \cref{alg:qpmcmc2_iter} commences by initializing these quantum registers to hold zero and subsequently executing five steps.

Initially, \cref{alg:qpmcmc2_iter} encodes the initial Markov chain state $\bt_{0}$
into the register $\mathcal{H}_0$. This operation necessitates approximately $\mathcal{O}(\log(|\mathcal{A}|))$ controlled-NOT gate operations where $\mathcal{A}$ is the parameter space (introduced in \cref{sec:MCMC and Barker's algorithm}).

Secondly, \cref{alg:qpmcmc2_iter} considers an operator $O_{\bar{q}}$ characterized by the Tjelmeland distribution $\bar{q}(\bt_{0},\cdot)$. This operation selects a state $\bar{\bt}$ from the distribution $\bar{q}(\bt_{0},\cdot)$ and encodes this state into the register $\mathcal{H}_1$. The resulting state is represented as:
$$
\ket{0}_{\mathcal{P}} \ket{\bt_0}_{\mathcal{H}_0}\ket{\bar{\bt}}_{\mathcal{H}_1}\ket{0}_{\mathcal{H}_2} \ket{0}_{\Pi}\ket{0}_{S},
$$
where $\bar{\bt}\sim \bar{q}\left(\bt_0, \cdot\right)$. The time required for this step is $\mathbb{T}(O_{\bar{q}})$.

Thirdly, \cref{alg:qpmcmc2_iter} creates a uniformly distributed superposition in register $\mathcal{P}$ such that each state is entangled with the proposal states $\bt_{p}$ encoded in register $\mathcal{H}_{2}$. This process can be achieved by employing approximately $\mathcal{O}(\log(P))$ rotation gate operations on register $\mathcal{P}$, followed by an operation $O_{\bar{q}}$ controlled by each $\ket{p}_{\mathcal{P}}$.
The resultant state is given by:
$$
\frac{1}{\sqrt{P+1}}\sum_{p=0}^{P}\ket{p}_{\mathcal{P}} \ket{\bt_0}_{\mathcal{H}_0}\ket{\bar{\bt}}_{\mathcal{H}_1}\ket{\bt_{p}}_{\mathcal{H}_2} \ket{0}_{\Pi}\ket{0}_{S},
$$
where $\bt_1, \ldots, \bt_P \stackrel{i.i.d.}{\sim} \bar{q}(\bar{\theta}, \cdot)$. Note that the time complexity for this operation is $\mathbb{T}(O_{\bar{q}}) + \mathcal{O}(1)$.

The fourth step involves encoding the evaluated value from the target distribution $\pi(\cdot)$ for each proposal label into the prefactor of each state. This task comprises two operations: the first is an oracle $O_{\pi_{\bar{\bt}}^*}$ that accepts queries from $\mathcal{H}_2$ and responds with the answer in the $\Pi$ register. Subsequently, a controlled rotation operator $CR$ receives a query from $\Pi$ and rotates the qubit in $\mathcal{S}$.
The resulting state is expressed as:

\begin{equation}
    \begin{aligned}
        & \frac{1}{\sqrt{P+1}}\sum_{p=0}^{P}\ket{p}_{\mathcal{P}} \ket{\bt_0}_{\mathcal{H}_0}\ket{\bar{\bt}}_{\mathcal{H}_1}\ket{\bt_{p}}_{\mathcal{H}_2} \ket{\pi_{\bar{\bt}}^*(\bt_{p})}_{\Pi} \left[ \sqrt{1-\pi_{\bar{\bt}}^*(\bt_p)}\ket{0}_{\mathcal{S}}+\sqrt{\pi_{\bar{\bt}}^*(\bt_p)}\ket{1}_{\mathcal{S}} \right] \\
        &=\sqrt{R}\ket{\text{SUCC}}_{\mathcal{P},\mathcal{H}_0,\mathcal{H}_1,\mathcal{H}_2,\Pi}\ket{1}_{\mathcal{S}}+\sqrt{1-R}\ket{\text{FAIL}}_{\mathcal{P},\mathcal{H}_0,\mathcal{H}_1,\mathcal{H}_2,\Pi}\ket{0}_{\mathcal{S}},
\end{aligned}\label{eq:final_state}
\end{equation}
where we denote $R=\frac{\sum_{p'=0}^{P}\pi_{\bar{\bt}}^*(\bt_{p'})}{P+1}$ and set $\ket{\text{SUCC}}$\footnote{To reduce the burden of notation, we omit the subscript.} as follows:
$$
\ket{\text{SUCC}}=  \sum_{p=0}^{P}
\sqrt{\frac{\pi_{\bar{\bt}}^*(\bt_{p})}{\sum_{p'=0}^{P}\pi_{\bar{\bt}}^*(\bt_{p'}) }}
\ket{p}_{\mathcal{P}}\ket{\bt_0}_{\mathcal{H}_0}\ket{\bar{\bt}}_{\mathcal{H}_1}\ket{\bt_{p}}_{\mathcal{H}_2} \ket{\pi_{\bar{\bt}}^*(\bt_{p})}_{\Pi}.
$$ 
The remaining states are left as $\sqrt{1-R}\ket{\text{FAIL}}$. Notice that for all $p=0,\ldots, P$, $\pi_{\bar{\bt}}^*(\bt_p) = \frac{\pi(\cdot)}{\pi(\bar{\bt})\mathcal{L}} \in [0,1]$. 
This guarantees that $\sqrt{1-\pi^*_{\bar{\bt}}}$ and $\sqrt{\pi^*_{\bar{\bt}}} \in [0,1]$. The time complexity of this task is $\mathbb{T}(O_{\pi_{\bar{\bt}}^*}) + \mathcal{O}(1)$.

In the final step, \cref{alg:qpmcmc2_iter} executes two measurements: the initial measurement targets the $\mathcal{S}$ register, followed by a subsequent measurement on the $\mathcal{H}_{2}$ register.
Should the qubit within the $\mathcal{S}$ register yield a state of $1$, the resultant state is altered to:
\begin{align*}
\ket{\text{SUCC}}
= 
\sum_{p=0}^{P}
\sqrt{\frac{\pi_{\bar{\bt}}^*(\bt_{p})}{\sum_{p'=0}^{P}\pi_{\bar{\bt}}^*(\bt_{p'}) }}
\ket{p}_{\mathcal{P}} \ket{\bt_0}_{\mathcal{H}_0}\ket{\bar{\bt}}_{\mathcal{H}_1}\ket{\bt_{p}}_{\mathcal{H}_2} \ket{\pi_{\bar{\bt}}^*(\bt_{p})}_{\Pi}\ket{1}_{\mathcal{S}}.
\end{align*}
Subsequently, \cref{alg:qpmcmc2_iter} performs a measurement on the $\mathcal{H}_{2}$ register, denoting the outcome as $\bt^{(s)}$, representing the selected state in the $s^{\text{th}}$ Markov chain.

Next, we give a proof of \cref{theorem:main}.

\begin{proof}
 The success rate $R$ of  \cref{alg:qpmcmc2_iter} is the probability of the event that the measurement in the register $\mathcal{S}$ yields $1$. According to \cref{eq:final_state}, $R$ has the following expression:
 \begin{equation}\label{success rate}
    R= \frac{\sum_{p'=0}^{P}\pi_{\bar{\bt}}^*(\bt_{p'})}{P+1}\ge\min_{p \in \{0,\ldots,P\}}(\pi_{\bar{\bt}}^*(\ttheta_p)).
\end{equation}

According to \cref{alg:qpmcmc2_iter}, with the generated intermediate state $\bar{\bt}$ and the given $\mathcal{L} \ge \max_{\bt \sim \bar{q}(\bt',\cdot ) ;\bt'\in\mathcal{A}}[\frac{\pi(\bt)}{\pi(\bar{\bt})}]$,
 we set $\pi_{\bar{\ttheta}}^*(\cdot)=\frac{\pi(\cdot)}{\pi(\bar{\ttheta}) \mathcal{L}}$.

 $$
 R\ge\min_{p \in \{0,\ldots,P\}}(\pi_{\bar{\bt}}^*(\ttheta_p))\ge 
 \frac{\min_{\bt \sim \bar{q}(\bt',\cdot ) ;\bt'\in\mathcal{A}}[\frac{\pi(\bt)}{\pi(\bar{\bt})}]}{\mathcal{L}}.
$$
  Consequently, \cref{theorem:main} follows.


\end{proof}
%





 
In the Section 3.3, we introduce an advanced version of \cref{alg:qpmcmc2_iter} with improved success rate using quantum amplitude amplification \citep{Brassard:2000xvp}.

\subsection{QPMCMC2 with Amplitude Amplification}\label{sec:AA}

Let $\chi: X \to \{0, 1\}$ be a Boolean function that partitions the set $X$ into “good” elements (where $\chi(x) = 1$) and “bad” elements. Consider a quantum algorithm $\mathbf{A}$ such that $\mathbf{A}|0\rangle = \sum_{x \in X} \alpha_x |x\rangle$, with  $a = \sum_{x \text{ is good}} |\alpha_x|^2$  representing the probability of producing a good element. If we repeatedly run $\mathcal{A}$, measure the output, and use $\chi$ to verify the result, the expected number of trials to find a good element is $1/a$.

Quantum amplitude amplification (AA) \citep{Brassard:2000xvp} is a well-studied technique that boost the amplitude of target state among superpositions. 
Using AA, the number of applications of $\mathbf{A}$  and its inverse needed to find a good element, without intermediate measurements, reduces to $\mathcal{O}(1/\sqrt{a})$. This generalizes Grover’s search  algorithm \citep{grover1996fast} and applies even when there is no promise of a unique solution.

Referring to \cref{alg:qpmcmc2_iter}, which maps $\ket{0}$ to \cref{eq:final_state} as follows: 
\begin{align}
\sqrt{R}\ket{\text{SUCC}}\ket{1}_{\mathcal{S}} + \sqrt{1-R}\ket{\text{FAIL}}\ket{0}_{\mathcal{S}}.
\end{align}
Applying quantum amplitude amplification with the setting where $\mathbf{A}$ is as defined in \cref{alg:qpmcmc2_iter}, and the ``good'' state corresponds to the superposition state associated with \( \ket{1}_{\mathcal{S}} \), the probability of measuring the good state is guaranteed to surpass $1/2$ after $\mathcal{O}(1/\sqrt{R})$ applications of lines 1-8 in \cref{alg:qpmcmc2_iter}. Thus, we have the \cref{cor:AA}.

In Section 4 and 5, we'll discuss the performance of \QPP \ through the implementation on inferring traits on a phylogenetic network.

\section{Case Study: Inferring Traits On a Phylogenetic Network}\label{sec:caseStudy}

In order to have a clearer image of how it works,
 we introduce the problem of inferring traits on a phylogenetic network as a suitable case study. In this section, we'll give brief introduction on this problem, then go through specific settings of  \cref{alg:qpmcmc2} we used in this case. Lastly, we'll provide further analysis on the success rate $R$ corresponding to our specific settings. The implementation results are left to section 5.
 
\subsection{Introduction to comparative phylogenetics and ancestral trait reconstruction}\label{sec:phylo}

\begin{figure}[!t]
	\centering
\includegraphics[width=0.9\textwidth]{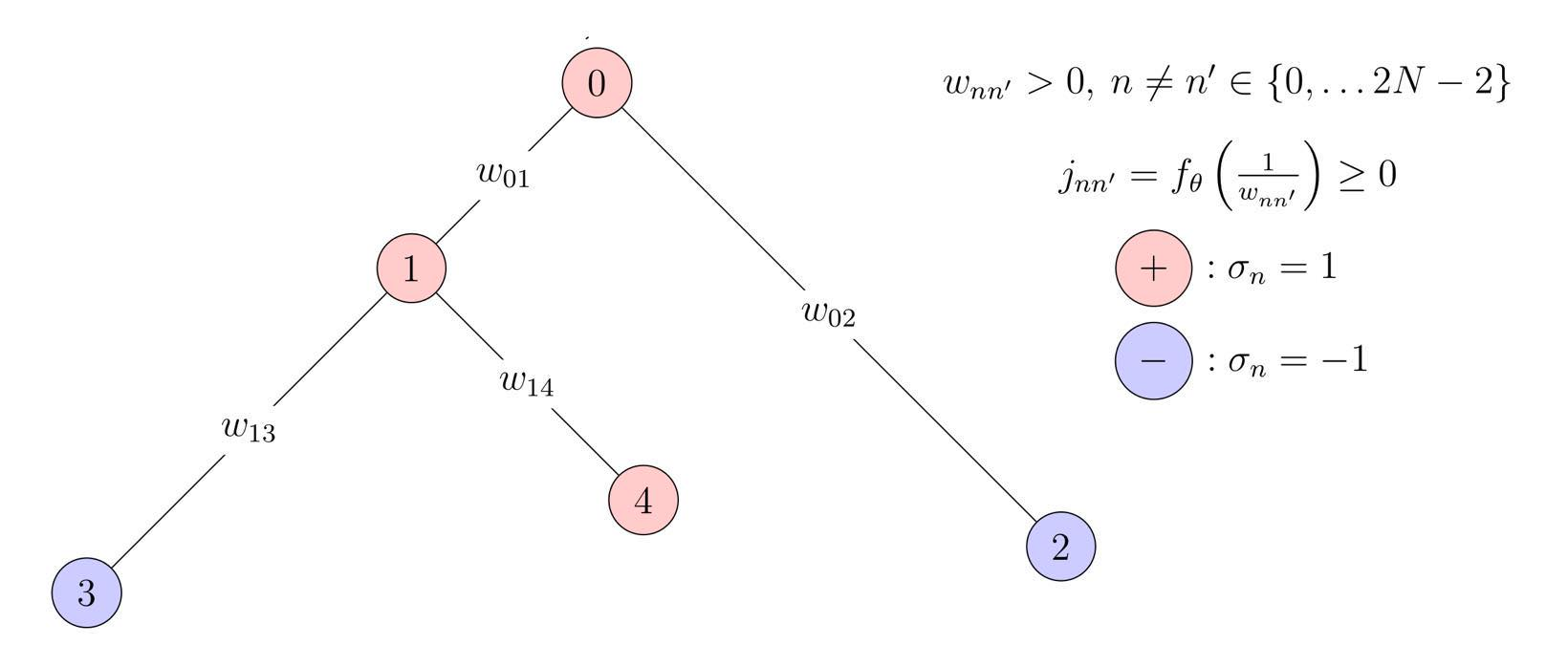}
\caption{This phylogenetic tree $\mathcal{G}$ has $M_o=3$ leaf nodes, $M_o-1=2=M_a$ internal nodes and $2M_o-1=5=M_{tot}$ total nodes. Leaf nodes represent observed taxa, and internal nodes are unobserved ancestors. We observe a binary trait variable $\sigma_m$ for each of the leaf nodes and model all (both observed and unobserved) traits $\sigma_m$ using an Ising model with interactions $j_{mm'}$ which condition on weights $w_{mm'}>0$.}\label{fig:phylo}
\end{figure}

Sampling algorithms are essential to the field of comparative phylogenetics, in general, and Bayesian phylogenetics \citep{suchard2018bayesian}, in particular.  Here, we start with a fixed phylogenetic tree structure and the traits of observed biological specimens (Figure \ref{fig:phylo}). We make the basic assumption that closely related taxa tend to share the same traits and establish a phylogenetic Ising model to predict the trait combinations of unbserved ancestors.  We also adapt this model to deviations from the basic tree graph structure in the context of bacterial reticulate evolution and extend this model to incorporate multiple traits. 

\newcommand{\ssigma}{\boldsymbol{\sigma}}

Specifically, suppose we assume a phylogenetic tree \( \mathcal{G} \) (Fig \ref{fig:phylo}) structure that describes the shared evolutionary history giving rise to \( M_o \) observed taxa indexed \( m \in \{M_o-1, \dots, 2M_o - 1=M_\text{tot}\} \). This phylogenetic tree is a rooted, undirected, bifurcating and weighted graph that contains \( M_\text{tot}=2M_o - 1 \) nodes, \( M_o \) of which (corresponding to observed taxa) are leaf nodes, and \( M_a=M_o - 1 \) are internal nodes. This graph also contains \( 2M_o - 2 \) edges, each of which has its own weight \( w_{mm'} > 0 \). If no edge exists between the node pair \( m, m' \), we say \( w_{m,m'} = \infty \). When edges exist, these weights are roughly proportional to the length of time spanning the existence of two organisms. Furthermore, suppose that we observe a binary trait, \( \sigma_m \in \{-1, 1\} \) for each of our observed taxa. We then may use a simple Ising model \citep{daskalakis2011evolutionary} to describe the joint distribution over observed and unobserved traits \( \ssigma = (\sigma_0, \dots, \sigma_{M_\text{tot}-1}) \):
\begin{equation}\label{eq:traits_posibility}
Pr(\boldsymbol{\sigma}|\beta,\gamma, \mathcal{G}) \propto \exp \left( \beta \sum_{m,m'} j_{mm'}\sigma_m \sigma_{m'} \right), \quad \mbox{where} \quad j_{mm'}=f_{\gamma}\left(\frac{1}{w_{mm'}}\right)
\end{equation}
and \( \beta > 0 \), \( f_\gamma : [0, \infty) \rightarrow [0,\infty) \), \( f_{\gamma}(0) = 0 \) and \( f_\gamma \) is an increasing function. For example, \( f_{\gamma}(x) = \gamma \sqrt{x} \) for \( \gamma > 0 \) is one of many possibilities. In the following, we treat \( \gamma \) and \( \beta \) as fixed, but one may learn them simultaneously with the rest of the model parameters in the context of Bayesian inference.
From (\ref{eq:traits_posibility}), we obtain the likelihood for the observed traits \( \ssigma_o = (\sigma_{M_a} , \dots , \sigma_{M_\text{tot}-1}) \) by conditioning on unobserved ancestral traits \( \ssigma_a = (\sigma_0, \dots , \sigma_{M_a-1}) \):
\begin{equation}\label{eq:likelihood}
Pr(\ssigma_o|\ssigma_a, \beta, \gamma, \mathcal{G}) \propto \exp \left( \beta \sum_{m,m'} j_{m,m'}\sigma_m \sigma_{m'} \right) .
\end{equation}
Placing the uniform prior on the ancestral traits $Pr(\ssigma_a) \propto 1$,
the posterior distribution for ancestral traits conditioned on observed traits becomes
\begin{equation}\label{eq:post_distribution}
 Pr(\ssigma_a|\ssigma_o, \beta, \gamma, \mathcal{G}) \propto Pr(\ssigma_o|\ssigma_a, \beta, \gamma, \mathcal{G}) \cdot Pr(\ssigma_a) \propto \exp \left( \beta \sum_{m,m'} j_{m,m'}\sigma_m \sigma_{m'} \right).
\end{equation}
Within the Bayesian paradigm of statistical inference, the problem of inferring unobserved ancestral traits \( \ssigma_a \) reduces to simulating from the Ising model (3) while keeping observed traits \( \ssigma_o \) fixed. Note that it is relatively simple to infer the joint posterior \( p(\ssigma_a, \beta, \gamma|\ssigma_o, \mathcal{G}) \), although we do not consider this task here.

\begin{figure}[h]
	\centering
 \includegraphics[width=0.7\textwidth]{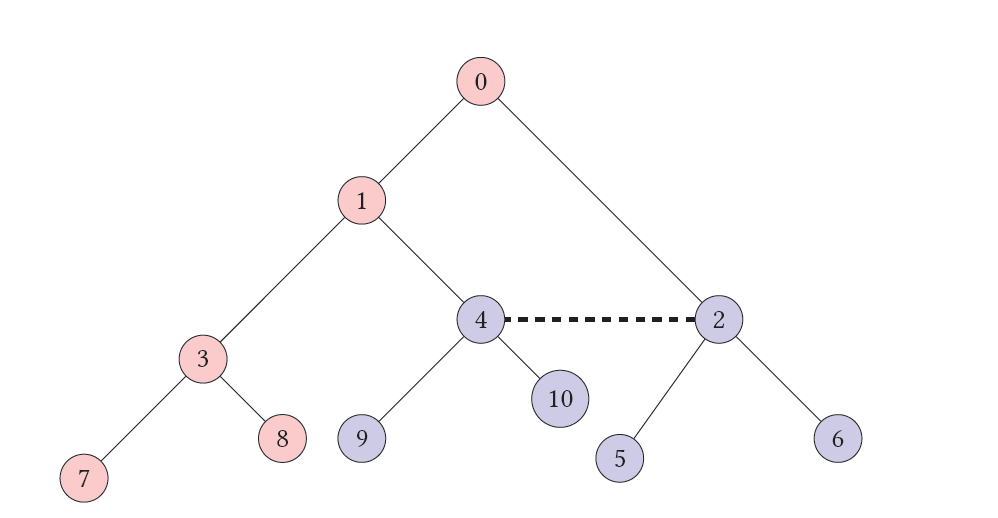}
\caption{Reticulate evolution.  This stylized bacterial phylogenetic network includes a reticulation (dashed line) that characterizes the exchange of genetic material between microbes. Whereas the network deviates from the bifurcating tree hypothesis of Figure \ref{fig:phylo}, the problem of ancestral trait reconstruction is still meaningful.}\label{fig:tree2}
\end{figure}

We build on this core model in two orthogonal ways. First, we consider the multi-trait scenario and model $T$ binary traits by allotting the $m^{\text{th}}$ specimen a spin of the form $\ssigma_m =(\sigma_{m,1},\dots,\sigma_{m,T})$. Following a development analogous to that of \eqref{eq:traits_posibility}, \eqref{eq:likelihood} and \eqref{eq:post_distribution}, we specify a multi-trait phylogenetic Ising model that leads to the posterior distribution
\begin{equation}\label{eq:potts_post}
Pr(\ssigma_a|\ssigma_o, \beta, \gamma, \mathcal{G}) \propto \exp \left( \beta \sum_{m,m'} j_{m,m'}\ssigma_m \cdot \ssigma_{m'} \right),
\end{equation}
and $\ssigma_a=(\ssigma_0,\dots,\ssigma_{M_a-1})$.
Second, we consider failures of the bifurcating evolutionary tree hypothesis.  Bacterial reticulate evolution (Figure \ref{fig:tree2}) arises from the exchange of genetic material between microbes.  In this context, it is appropriate to model evolution using a phylogenetic network.  The Neighbor-net \citep{neighborNet} algorithm is a popular algorithm for phylogenetic network construction that uses distances between genetic sequences to construct a planar splits graph. In this graph, extremal nodes are observed specimens, and interior nodes are potential ancestors. Whereas this evolutionary network model does not represent an explicit history of individual reticulations, it does represent conflicting signals regarding potential reticulations.  These candidate reticulations take the form of the interior boxes that manifest in Figure \ref{fig:bactNet}.

On the one hand, using such a phylogenetic network as the base lattice structure in phylogenetic Ising models does not alter the mathematical details of the posterior distributions \eqref{eq:post_distribution} and \eqref{eq:potts_post}.  On the other hand, the existence of cycles in the splits graph makes sampling these distributions significantly more difficult.  \cite{holbrook2023quantum} shows Algorithm \ref{alg:multiProp}'s potential for sampling from such challenging target distributions and advances \QP, which approximately performs this algorithm. In the following, we present  \QPP\ and its massive speedups over the $\mathcal{O}(P)$ complexity of Algorithm \ref{alg:multiProp} and the $\mathcal{O}(\sqrt{P})$ complexity of \QP.

\subsection{Functions used in QPMCMC2}
The learning of ancestral traits (Section \ref{sec:phylo}) within a known phylogenetic network illustrates our algorithm's speed, flexibility and fully-explicit nature. Consider a phylogenetic network $\mathcal{G}(V,E)$, where $V$ denotes a set of $M_\text{tot}$ vertices and $E$ represents a set of edges.
Let $V_{o}$ be a designated subset of $V$ signifying the observed taxa within this context, and $V_{a}=V\backslash V_{o}$ be the complement set of $V_{o}$. 
For the network shown in Figure \ref{fig:tree2}, we have $M_{tot}=|V|=11$, $V_{o}=\{5,\dots,10\}$ and $V_{a}=\{0,\dots,4\}$.
Using the notation of Sections \ref{sec:MCMC and Barker's algorithm}, \ref{sec:multipropMCMC} and \ref{sec:main}, we identify any Markov chain state with a collection of ancestral traits thus:
\begin{equation}\label{eq:genericState}
\boldsymbol{\theta} = (\boldsymbol{\sigma}_0,\boldsymbol{\sigma}_1, \ldots, \boldsymbol{\sigma}_{{M_a}-1})\,, 
\end{equation}
where  $\boldsymbol{\sigma}_{m} = (\sigma_{m,1}, \ldots, \sigma_{m,T})$ for $\sigma_{m,t}\in\{-1,1\}$. 
Assuming that each \( (m,m') \in E \) possesses an identical weight $J$, we rewrite the posterior \eqref{eq:potts_post} as
\begin{align}\label{eq:pi for Ising model}
Pr(\ssigma_a|\ssigma_o,J,\mathcal{G}) \propto \exp\left(J \sum_{(m,m') \in E} \boldsymbol{\sigma}_{m} \cdot \boldsymbol{\sigma}_{m'}\right) \quad \mbox{and set} \quad \pi(\ttheta) := Pr(\ssigma_a|\ssigma_o,J,\mathcal{G}) \, .
\end{align}
 Fixing the observed traits and sampling unobserved ancestral traits using \QPP\ amounts to efficient posterior inference.

In the following, we specify the Tjelmeland distribution $\bar{q}(\cdot,\cdot)$ and detail the target distribution $\pi^{*}(\cdot)$ for the phylogenetic Ising model. 
Next, we analyze the qubit requirements when applying \cref{alg:qpmcmc2} for this specific inferential task. 
Finally, the success rate $R$ of \cref{alg:qpmcmc2} in this application scenario is introduced.

\subsubsection{Tjelmeland distribution \texorpdfstring{$\bar{q}(\cdot,\cdot)$}{q}}\label{sec:4_1}

We specify the symmetric Tjelmeland distribution $\bar{q}(\cdot,\cdot)$ by defining the distribution $\bar{q}(\ttheta,\cdot)$ centered at a generic state \eqref{eq:genericState}.
For each $t \in \{1,\dots,T\}=\mathcal{T}$ and $m \in \{0,\dots,M_a-1\}=V_a$, we define the result state
\begin{equation}
\boldsymbol{\theta}_{m,t}=(\boldsymbol{\sigma}_0,\ldots \boldsymbol{\sigma}_m',\ldots,\boldsymbol{\sigma}_{M_{a}-1}),
\end{equation}
where 
$\boldsymbol{\sigma}_m'=(\sigma_{m,1},\ldots -\sigma_{m,t},\ldots,\sigma_{m,T})$. The vectors $\boldsymbol{\theta}$ and $\boldsymbol{\theta}_{m,t}$ only differ by a negative sign at the trait $(t,m)$.  
Since there are $M_a T=(M_\text{tot}-M_o)T$ possibilities of $\bt_{m,t}$, 
we write down $\bar{q}(\boldsymbol{\theta},\boldsymbol{\theta}')$ formally as 
\begin{equation}\label{qbar_Ising}
 \bar{q}(\boldsymbol{\theta},\boldsymbol{\theta}')=
 \begin{cases}
    \frac{1}{M_a T+1}& \text{if } \boldsymbol{\theta}' \in \boldsymbol{\Theta} \\
    0              & \text{otherwise} \, ,
\end{cases}
\end{equation}
where $\boldsymbol{\Theta}=
\{\bt_{m,t}: t\in \mathcal{T} \text{ and } m \in V_a   \} \cup \{\boldsymbol{\theta}\}$.  
In words, $\bar{q}(\ttheta,\cdot)$ is a uniform distribution over the nearest neighbors to $\ttheta$ and $\ttheta$ itself.

Notice that we are able to provide $L$ with this
this given form Tjelmeland distribution $\bar{q}(\cdot,\cdot)$. For two states $\bar{\ttheta}$ and $\ttheta\sim\bar{q}(\bar{\ttheta},\cdot)$, they only differ in at most one bit. This leads to the existance of $\mathcal{L}$:

\begin{equation}\label{eq:q_condition_ising}
    0<e^{-2J\text{deg}(\mathcal{G})}\le \frac{\pi(\ttheta)}{\pi(\bar{\ttheta})}\le e^{2J\text{deg}(\mathcal{G})}=:\mathcal{L}.
\end{equation}

Here, we find the value of $\mathcal{M}:=\min_{\bt \sim \bar{q}(\bt',\cdot ) ;\bt'\in\mathcal{A}}[\frac{\pi(\bt)}{\pi(\bt')}]=e^{-2J\text{deg}(\mathcal{G})}$ described in \cref{theorem:main}, too.
According to \cref{theorem:main}, using this Tjelmeland distribution $\bar{q}(\cdot,\cdot)$ \QPP\ are able to run with time complexity independent of $P$. The success rate of \cref{alg:qpmcmc2_iter} is lowerbounded as follows:
 
\begin{equation} \label{eq:R_lowerbound_Ising}
    1\ge R\ge \frac{\mathcal{M}}{\mathcal{L}}=e^{-4J\text{deg}(\mathcal{G})}
\end{equation}

With \cref{eq:R_lowerbound_Ising}, we analyze how graph types influence the success rate \(R\) of \cref{alg:qpmcmc2_iter}: 
for graphs such as ideal tree graphs and 2D square lattice graphs, \(\text{deg}(\mathcal{G})\) is guaranteed to be small, resulting in a higher success rate \(R\). In contrast, graphs like star graphs can exhibit very high degrees \(\text{deg}(\mathcal{G})\) depend on the number of "legs", making the success rate \(R\) exponentially small. As a result, our proposed method \QPP\ tends to be less efficient in scenarios where \(\text{deg}(\mathcal{G})\) is large. 

Fortunately, for the ancestral trait reconstruction problems we are interested in, \(\text{deg}(\mathcal{G})\) is generally small: for an ideal tree, \(\text{deg}(\mathcal{G}) = 3\). The \(\text{deg}(\mathcal{G})\) of a realistic phylogenetic tree could exceed $3$ due to the reticular evolution, however, the degrees remain small in general.
In \cref{sec: experiments}, we focus on sampling Ising models from a 2D square lattice graph and a realistic \textit{Salmonella} phylogenetic tree with \(\text{deg}(\mathcal{G}) = 8\). With small given $J$s, in these cases, \QPP\ demonstrates high efficiency with high success rate \(R\)s.

\subsubsection{Target function \texorpdfstring{$\pi_{\bar{\bt}}^*(\cdot)$}{pi}}
\label{sec:4_2}
To introduce the relative target distribution $\pi_{\bar{\bt}}^*(\cdot)$ in \cref{alg:qpmcmc2}, we first define a function $f_{(m,t)}$ which maps a  state $\bt$ in parameter space to $\mathbb{R}^{+}$ as follows:
$$
f_{(m,t)}({\boldsymbol{\theta}})=
\sum_{m';(m',m)\in E} \sigma_{m,t}\cdot \sigma_{m',t}+\text{deg}(\mathcal{G}), $$
where ${\sigma}_{m,t}$ is the trait of $\bt$ and $\text{deg}(\mathcal{G})$ is the degree of the phylogenetic network $\mathcal{G}$. 

Considering the Tjelmeland distribution \eqref{qbar_Ising} in \cref{alg:qpmcmc2}, each proposal state $\boldsymbol{\theta}_p$ has at most one trait that is different from the intermediate state $\bar{\boldsymbol{\theta}}$. 
Therefore, the function $\pi_{\bar{\bt}}^*(\cdot)$  in \cref{alg:qpmcmc2} that satisfied $\pi_{\bar{\bt}}^*(\cdot) \propto \pi(\cdot)$ can be expressed as 
\begin{align}\label{eq.pi_star}
\pi_{\bar{\bt}}^*(\ttheta_p;\bar{\bt}) = 
\left\{
\begin{array}{cc}
    \exp{-2Jf_{(m_{p},t_{p})}(\bar{\bt})} & \text{if }\boldsymbol{\theta}_p\neq \bar{\boldsymbol{\theta}}\\
    \exp{-2J\text{deg}(\mathcal{G})} & \text{otherwise}\, ,
\end{array}
\right.
\end{align}
where $(m_p,t_p)$ is the flipped trait in the proposal state $\mathbf{\theta}_{p}$. Note that the image of the function $\pi_{\bar{\bt}}^*(\cdot)$ belongs to $(0,1]$.

\subsubsection{Qubit requirement}

Given the Tjelmeland distribution \cref{qbar_Ising} and the  target function \cref{eq.pi_star} introduced in this section, we can analyze the qubit requirement for \cref{alg:qpmcmc2}. Encoding $\bt_{p}$ in $\mathcal{H}_2$ requires $\lceil TM_{a} \log_2(TM_a) \rceil$ qubits, which becomes infeasible for near-term applications. However, this dilemma can be mitigated by encoding $(m_p,t_p)$ (the flipped trait in the proposal state $\bt_{p}$), which is sufficient for calculating the relative target distribution $\pi_{\bar{\bt}}^*(\cdot)$ and requires only $\lceil \log_2(TM_a) \rceil$ qubits.

Secondly, the calculation of the function $\pi_{\bar{\bt}}^*(\cdot)$ is required for each iteration in \cref{alg:qpmcmc2}, which is relatively challenging for a near-term quantum computer due to the complexity of computing this exponential function. However, by considering a constant $J$ in \cref{eq.pi_star}, we can pre-calculate $2\text{deg}(\mathcal{G})+1$ possibilities of the image of \cref{eq.pi_star}. Consequently, the calculation of $\pi_{\bar{\bt}}^*(\bt_{p})$ can be obtained by providing $(m_p,t_p)$ and consulting a lookup table.

\subsection{Success rate analysis}

In this subsection, instead of deriving the lower bound as in \cref{eq:R_lowerbound_Ising}, we focus on the expectation value of the success rate $R$ over different random proposal sets. We believe this serves as a better benchmark for evaluating the efficiency of our algorithm.

This subsection comprises two aspects. First, for a given input state $\ttheta_0$ and a intermediate state $\bar{\ttheta}$ selected according to $\bar{q}(\ttheta_0,\cdot)$,  we introduce the expectation and variance of $R$
in \cref{th.Ising-acceptance rate} in terms of $\ttheta_0$, $\bar{\ttheta}$ and some problem-dependent parameters, over all possible sets of proposals generated according to $\bar{q}(\bar{\ttheta},\cdot)$.
Second, we provide the proof of \cref{th.Ising-acceptance rate}.

\begin{theorem} \label{th.Ising-acceptance rate}
    Consider one iteration in \cref{alg:qpmcmc2} by providing the number of proposals $P$, the previous state $\bt_0$, the relative target distribution $\pi^{*}$ defined in \cref{eq.pi_star}, and the Tjelmeland distribution defined in \cref{qbar_Ising}. For a given Ising distribution with the coupling constant $J$ and graph $\mathcal{G}$, the expectation $\mathbb{E}[R]$ and variance $\mathbb{V}[R]$ of $R$ over all possible proposal sets \( \{\bt_0, \ldots, \bt_P\} \), where \( \bt_{p} \stackrel{iid}{\sim} \bar{q}(\bar{\ttheta}, \cdot) \) for $p = 1, \ldots, P $, can be bounded as follows:
\begin{align}
\mathbb{E}[R]
&\geq Pr(\bt_0)^{\frac{-4}{TM_a}}
\exp[
-2J
\text{deg}(\mathcal{G})
(1+\epsilon_2+\frac{4}{TM_a})]
(1-\epsilon_1)
+
\exp[-2J f_{(m_0,t_0)}(\bar{\bt})]\epsilon_1.
\\
\mathbb{V}[R]
&\leq \epsilon_1.
\end{align}
Here $(m_0,t_0)$ is the flipped trait in the previous state $\bt_{0}$. 
We denote $\epsilon_1=\frac{1}{P+1}$,  $\epsilon_2=\frac{M_o}{M_{a}}$, and     $Pr(\ttheta)=\exp(J\sum_{t \in \mathcal{T}} \sum_{(m,m')\in E} \sigma_{m,t}\cdot \sigma_{m',t})$ where $\sigma_{m,t}$ are the traits of $\bt$ as expressed in \cref{eq:genericState}.

\end{theorem}

This proposition suggests that the expected success rate $R$ is approximated by  $\exp[-2J\text{deg}(\mathcal{G})]$, and the variance of $R$ approaches $0$ when $\epsilon_1$ and $\epsilon_2$ are close to $0$ when $TM_a$ is large.
To address $\epsilon_1$, we can consistently set it to a small value by increasing the number of proposals in \QPP.
Regarding $\epsilon_2$, we observe that a realistic phylogenetic network graph $\mathcal{G}$ may feature numerous reticulations and $M_a$ being much larger than $M_o$.

\begin{proof}
The success rate $R$ (defined in \cref{success rate}) has the following expression:
$$
R=\frac{\sum_{p=0}^{P}\pi_{\bar{\bt}}^*(\bt_{p})}{P+1}.
$$
Using \cref{eq.pi_star}, the expected value of $R$ denoted as $\mathbb{E}[R]$ is given by:

\begin{align*}
\mathbb{E}[R]
=
\frac{\sum_{t\in \mathcal{T}}\sum_{m \in V_a}
\exp(-2J f_{(m,t)}(\bar{\bt}))}{TM_a}
(1-\epsilon_1)
+
\exp(-2J f_{(m_0,t_0)}(\bar{\bt}))\epsilon_1,
\end{align*}
where $\frac{1}{P+1}$ is set as $\epsilon_1$.

The first term can be bounded using the AM-GM inequality as follows:
\begin{align}
&\frac{\sum_{t\in \mathcal{T}}\sum_{m \in V_a}
\exp(-2J f_{(m,t)}(\bar{\bt}))}{TM_a}
\geq
\exp( \frac{-2J}{TM_{a}}\sum_{t\in \mathcal{T}}\sum_{m \in V_a} f_{(m,t)}(\bar{\bt}))
\nonumber
\\
=
&
\exp( 
\frac{-2J}{TM_{a}}\sum_{t\in \mathcal{T}}\sum_{m \in V_a} 
\left(\sum_{m';(m',m)\in E} 
\bar{\sigma}_{m,t}\cdot \bar{\sigma}_{m',t}+\text{deg}(\mathcal{G})\right)
), \label{eq:AMGM}
\end{align}
and the exponent in the above quantity can be upper bounded as follows: 

\begin{align*}
&\sum_{t\in \mathcal{T}}\sum_{m \in V_a} 
\left( \sum_{m';(m',m)\in E} 
\bar{\sigma}_{m,t}\cdot \bar{\sigma}_{m',t}+\text{deg}(\mathcal{G})\right)
\\
=
&\sum_{t\in \mathcal{T}}\sum_{m \in V} 
\sum_{m';(m',m)\in E} 
(\bar{\sigma}_{m,t}\cdot \bar{\sigma}_{m',t})
-
\sum_{t\in \mathcal{T}}\sum_{m \in V_{o}} 
\sum_{m';(m',m)\in E} 
(\bar{\sigma}_{m,t}\cdot \bar{\sigma}_{m',t}) + TM_a \text{deg}(\mathcal{G}) 
\\
\leq
&
2
\sum_{t\in \mathcal{T}}
\sum_{(m',m)\in E} 
(\bar{\sigma}_{m,t}\cdot \bar{\sigma}_{m',t})+TM_{tot} \text{deg}(\mathcal{G}).
\end{align*}

The equality arises from the fact that $V_a=V\backslash V_o$ and $M_a=|V_a|$.
For the inequality, note that $\bar{\sigma}_{m,t}\cdot \bar{\sigma}_{m',t}\in \{-1,1\}$, which implies that the upper bound of the second term is $TM_{o}\text{deg}(\mathcal{G})$.
Therefore, \cref{eq:AMGM} can be upper bounded as follows:
\begin{align*}
&\exp( 
\frac{-4J}{TM_a}
\sum_{t\in \mathcal{T}}
\sum_{(m',m)\in E} 
(\bar{\sigma}_{m,t}\cdot \bar{\sigma}_{m',t})-2J\text{deg}(\mathcal{G})\frac{M_\text{tot}}{M_a})
\\
=&
Pr(\bar{\bt})^{\frac{-4}{TM_a}}
\exp[
-2J(\text{deg}(\mathcal{G})
)(1+\epsilon_2)],
\end{align*}
where we set $\frac{M_o}{M_a}$ as $\epsilon_2$, and set $Pr(\ttheta)=\exp(J\sum_{t \in \mathcal{T}} \sum_{(m,m')\in E} \sigma_{m,t}\cdot \sigma_{m',t})$. 

Furthermore, the Tjelmeland distribution introduced in \cref{sec:4_1} implies $Pr(\bar{\ttheta})\leq Pr(\ttheta_0)e^{2J\text{deg}(\mathcal{G})}$, we conclude the expectation of $R$ can be lower bounded as follows: 

\begin{align*}
& \mathbb{E}[R]\geq 
(1-\epsilon_1)
Pr(\bar{\bt})^{\frac{-4}{TM_a}}
\exp[
-2J(\text{deg}(\mathcal{G})
)(1+\epsilon_2)]
+
\exp[-2J f_{(m_0,t_0)}(\bar{\bt})]\epsilon_1
\\
\geq
&
(1-\epsilon_1)
Pr(\bt_0)^{\frac{-4}{TM_a}}
\exp[
-2J
(\text{deg}(\mathcal{G})
(1+\epsilon_2+\frac{4}{TM_a})]
+
\exp[-2J f_{(m_0,t_0)}(\bar{\bt})]\epsilon_1.
\end{align*}

For the variance of $R$, it can be upper bounded as follows:
$$
\mathbb{V}[R]
=
\mathbb{V}\left[\frac{\sum_{p=0}^{P}\pi_{\bar{\bt}}^*(\bt_{p})}{P+1}\right]
=
\frac{\mathbb{V}\left[\sum_{p=0}^{P}\pi_{\bar{\bt}}^*(\bt_{p})\right]}{(P+1)^2}
=
\frac{\mathbb{V}[\pi_{\bar{\bt}}^*(\bt_{p})]}{P+1}
\leq
\epsilon_1,
$$
where the inequality uses the the fact $\pi_{\bar{\bt}}^*(\bt_{p})\in[0,1]$ which implies $\mathbb{V}[\pi_{\bar{\bt}}^*(\bt_{p})]\leq 1$.

\end{proof}

\section{Application: Salmonella and Antibiotic Resistance} \label{sec: experiments}

\begin{figure}[!t]\label{fig:bactNet}
    \centering
    \hspace*{-2cm}
    \includegraphics[width=0.75\linewidth]{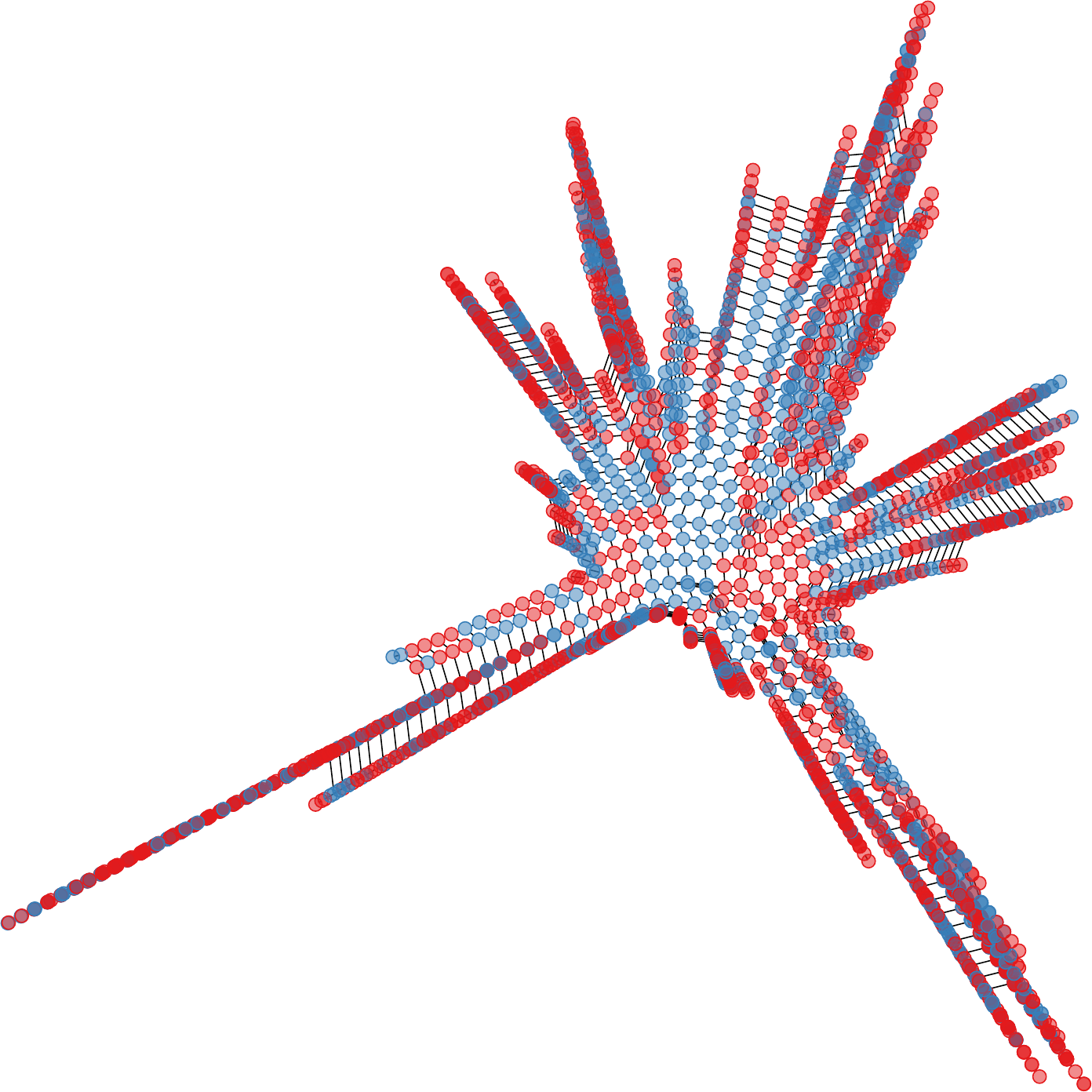}
    \label{fig:bactNet_Salmonella}

    \caption{
      A Neighbor-net phylogenetic network describes the shared evolutionary history of 248 salmonella bacteria isolates. The extremal nodes correspond to the 248 observed isolates, and the $M_a=3{,}065$ interior nodes correspond to unobserved ancestors. Interior squares are potential reticulation events. Colors (red, resistance; blue, no resistance) are observed and posterior mode resistances to the antibiotic ampicillin for observed microbes and unobserved ancestors, respectively.}
\end{figure}

Conditioned on antibacterial drug resistance scores for 248 \textit{Salmonella} bacterial isolates, we apply our \QPP\ algorithm to the Bayesian inference of ancestral traits on a Neighbor-net phylogenetic network (Section \ref{sec:phylo}). \citet{mather2013distinguishable,cybis2015assessing} previously used this biological dataset to analyze the development of antibiotic resistances within the genus \textit{Salmonella}, but their analyses did not account for bacterial reticulate evolution. Our phylogenetic network, denoted as \(\mathcal{G}_{\text{sal}}(V,E)\), comprises \(M_{\text{tot}}=3{,}313\) vertices and \(|E|=5{,}945\) edges. Among these vertices, there are \(M_{\text{o}}=248\) observed taxa, representing the observed biological isolates with known traits. Pertinent to the theoretical developments in Section \ref{sec: experiments}, the degree of our network is \(\text{deg}(\mathcal{G}_{\text{sal}})=8\). In this section, we use a classical simulator to execute \cref{alg:qpmcmc2} and evaluate its efficiency.  We apply our algorithm to two Neighbor-net phylogenetic networks: 1) a square lattice graph  which contains $100\times100$ interior nodes with additional 400 extremal nodes along 4 sides, representing the observed isolates, and 2) the aforementioned Neighbor-net phylogenetic network describing the shared evolutionary history of 248 \textit{Salmonella} bacterial isolates (see \cref{fig:bactNet}). Additionally, we consider two cases: the single-trait case with the trait number \(T=1\) and the multi-trait case with the trait number \(T=4\), accounting for four traits in each \textit{Salmonella} bacterial isolate. 
Each figure is plotted with results averaged over 10 repetitions of the experiment. 
The implementation code is available at \url{https://github.com/CYLin1113/Quantum-Parallel-MCMC-2}. To check the implementational correctness of our code, we run \QPP\ on a 3 times 3 square lattice model, see \cref{App. A}.

In all experiments presented in \cref{sec: experiments}, the coupling constants \(J\) of the Ising models are set to \(J=0.3\) for square lattices and \(J=0.03\) for the \textit{Salmonella} phylogenetic tree. These selections of \(J\) result in high success rates \(R\), and their impact on the running time of \QPP\ (\cref{alg:qpmcmc2}) can be mitigated by executing multiple copies of \cref{alg:qpmcmc2_iter} simultaneously, with additional qubits independent of proposal number $P$. Please note that in specific cases with large $J$s or large $\text{deg}(\mathcal{G})$s of the target phylogenetic trees, the acceptance rate $R$s could be very small, although they remain independent of $P$. In such cases, amplitude amplification techniques, as described in \cref{sec:AA}, can be applied to efficiently boost the success rate $R$ over $0.5$ \citep{Brassard:2000xvp}.

\begin{figure}[!t]   
    \centering

    \subfigure[Square lattices]{ \includegraphics[width=0.45\textwidth]{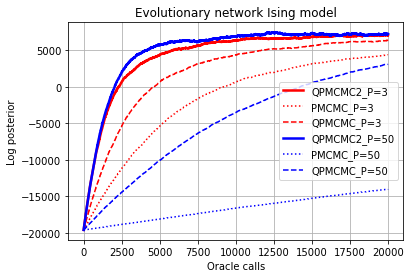}}
    \subfigure[Salmonella phylogenetic network]{\includegraphics[width=0.45\textwidth]{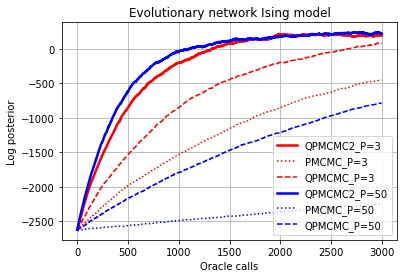}
     }

 \caption{Comparison between trace plots generated by the \QPP, \QP \citep{holbrook2023quantum} and \cref{alg:multiProp} (PMCMC) for $P=3$ and $P=50$. For implementation of one MCMC iteration, \QPP\ requires 1 oracle calls of $\pi_{\bar{\bt}}^*(\cdot)$, while \QP \ requires $\mathcal{O}(\sqrt{P})$ calls and multiproposal MCMC requires $P+1$ calls. In these cases, only \QPP\ improves the converge rate when using a larger $P$.}
    \label{fig: Time complexity comparison}
\end{figure}

From the relationships shown in \cref{fig: Time complexity comparison}, we demonstrate the value of \QPP\: the classical version of multiproposal MCMC \cref{alg:multiProp}, labeled as PMCMC, requires $\mathcal{O}(P)$ oracle calls to execute one iteration, which slows down the convergence process when using a larger number of proposals. With the help of quantum parallel computing in our approach, \QPP\ is able to compute all \(P\) proposals in parallel with $\mathcal{O}(\log P)$ qubits during each iteration. This resolves the computational bottleneck of using large \(P\) values in \cref{alg:multiProp}, where we find potential advantages exist.

\begin{figure}[!t]  
    \centering

    \subfigure[Single-trait (square lattices) ]{ \includegraphics[width=0.45\textwidth]{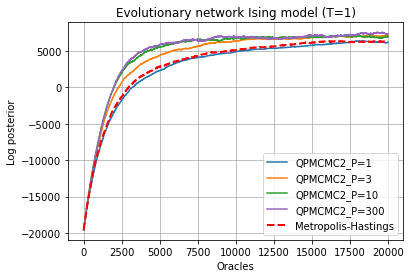}}
    \subfigure[Four-trait (square lattices) ]{ \includegraphics[width=0.45\textwidth]{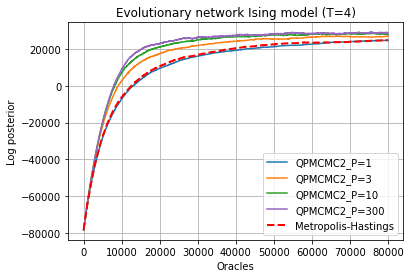}}
 \caption{Trace plots generated by the \QPP \ algorithm for different numbers of proposals $P$ and Metropolis-Hastings algorithm, tested on the square lattice graph. For both the single-trait and multi-trait problem, increasing $P$ accelerates convergence to higher posterior probability states. Here, we observe that \QPP\ significantly outperforms the Metropolis-Hastings algorithm: for $P=300$, \QPP\ achieves a 3.8-fold improvement in convergence rate compared to the Metropolis-Hastings algorithm.
 }
 
    \label{fig: Single traits (2D)}
\end{figure}

\begin{figure}[!t]   
    \centering

    \subfigure[Single-trait (salmonella)]{ \includegraphics[width=0.45\textwidth]{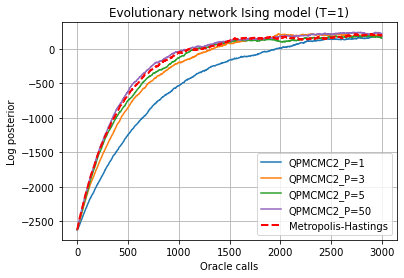}}
    \subfigure[Four-trait (salmonella)]{\includegraphics[width=0.45\textwidth]{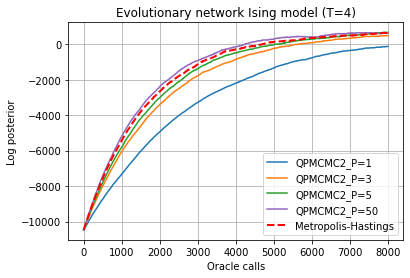}
     }

 \caption{Trace plots generated by the \QPP \ algorithm for different numbers of proposals $P$ and Metropolis-Hastings algorithm,  tested on the phylogenetic tree of salmonella bacteria isolates (\cref{fig:bactNet}). For both the single-trait and multi-trait problem, increasing $P$ accelerates convergence to higher posterior probability states. Due to the higher degree $\text{deg}(\mathcal{G}_{sal})=8$ which leads to a lower success rate (\cref{eq:R_lowerbound_Ising}),  we failed to observe the same relative performance gain over MH as we observed for the square lattices model. 
 }
 
    \label{fig: Single traits}
\end{figure}

In \cref{fig: Single traits (2D)} and \cref{fig: Single traits}, we include the Metropolis-Hastings (MH) algorithm in our analysis, as it is generally considered more efficient than Barker-acceptance-based MCMC. From the relationships shown in plot (a) of \cref{fig: Single traits (2D)} and \cref{fig: Single traits}, it is evident that by evaluating more proposals in a single iteration, \QPP\ converges faster and eventually surpasses the MH algorithm. In the case of the square lattice graph, this speedup is more significant: \QPP\ converges 3.8 times faster compared to the MH algorithm when \(P = 300\). These results indicate that, with the aid of quantum parallel computing, this Barker-acceptance-based multiproposal MCMC can approach or even surpass the efficiency of the Metropolis-Hastings algorithm.

We not only apply this method to cases with a single trait (\(T=1\)) but also extend \QPP\ to a phylogenetic network Ising model with multiple traits (ampicillin, chloramphenicol, ciprofloxacin, and furazolidone resistances). Plot (b) in \cref{fig: Single traits (2D)} and \cref{fig: Single traits} shows the trace plot for the corresponding log-posterior with \(T=4\). As the number of parallel proposals \(P\) increases, the ancestral trait configuration tends to converge faster while maintaining detailed balance. As expected, the algorithm appears to require approximately \(T\) times the number of iterations compared to the \(T=1\) case.

\begin{figure}[!t]   
    \centering
    
    \subfigure[Square lattices]{\includegraphics[width=0.45\textwidth]{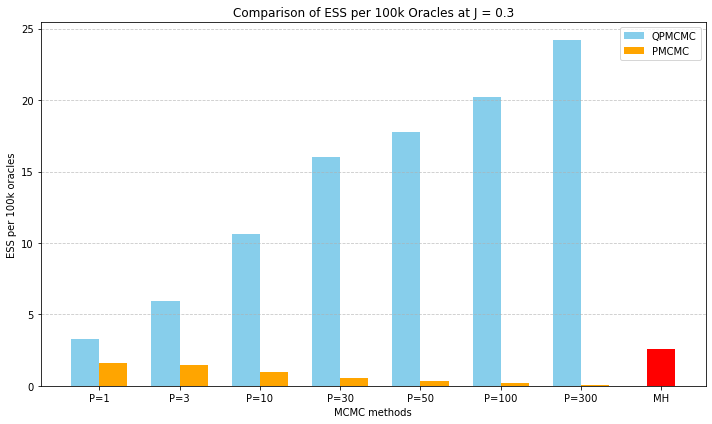}
     }
    \subfigure[Salmonella phylogenetic tree]{ \includegraphics[width=0.45\textwidth]{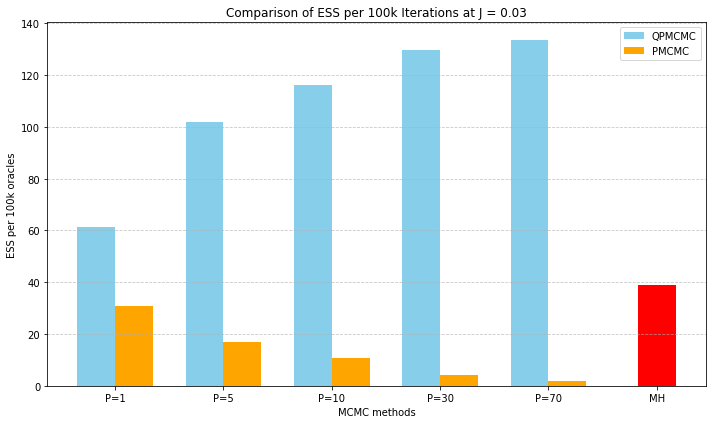} }

    \caption{Effective sample size (ESS) for the log posterior per 100,000 oracle calls for different numbers of proposals $P$. ESS produced by \QPP \ shows a significant improvement  as $P$ increases,  with a noticeable gap compared to the results of the MH algorithm: using \QPP, we obtain a $11$-fold advantage in the case of square lattices, while it's a $3.5$-fold advantage in the case of salmonella phylogenetic tree, comparing to the MH algorithm.
    In contrast,  classical multiproposal MCMC (PMCMC) exhibits the decreasing performance due to the $\mathcal{O}(P)$ complexity using $P$ proposals. } \label{fig: ESS}
\end{figure}

Next, we focus on comparing the ESS per oracle among Algorithm \ref{alg:multiProp} (PMCMC), Algorithm \ref{alg:qpmcmc2} (\QPP), and the Metropolis-Hastings (MH) algorithm. We estimate ESS using the \texttt{Python} package \texttt{ArviZ} \citep{kumar2019arviz}.  As shown in \cref{fig: ESS}, the ESS per 10k oracles increases significantly when a larger number of proposals \(P\) is used in \QPP. For the square lattice graph and \textit{Salmonella} phylogenetic tree, with sufficiently large \(P\), \QPP\ generates samples with ESS values that are 11 and 3.5 times greater than those produced by the MH algorithm, respectively, demonstrating the remarkable advantages of \QPP\ over this classical approach. The same improvement is not observed in the classical multiproposal MCMC (labeled as PMCMC) due to its \(\mathcal{O}(P)\) cost. 


In both the single- and the multi-trait experiments, we  observe that using a large proposal count \(P\) when applying multiproposal MCMC leads to improved convergence. This highlights two major strengths of the quantum algorithm we propose.
First, \QPP\ obtains an exponential speedup: for large \(P\) in multiproposal MCMC algorithms, we reduce the dependence of $P$ of the time complexity from \(\mathcal{O}(P)\) to \(\mathcal{O}(1)\) with \(\mathcal{O}(\log P)\) ancillary qubits. This is an exponential speedup as a function of $P$ and resolves the bottleneck of the original Algorithm \ref{alg:multiProp}. 
Second, \QPP\ provides accelerated sampling for real-world problems: we have demonstrate the benefits for our quantum algorithm in accelerating sampling for a realistic and non-trivial class of graphical models. This quantum algorithm shows the potential to accelerate Bayesian reconstruction of bacterial antibiotic resistances, an important problem in medicine and evolutionary biology.

\section{Discussion}

Quantum computing is set to revolutionize certain areas of science (computational physics/chemistry), but its future impact on many other areas remains unknown.  Similarly, quantum computing promises extreme speedups for certain technical challenges (prime factorization in cryptography) while benefits for other prominent challenges remain elusive.  In particular, many statisticians may wonder how quantum computing will eventually impact their day-to-day data scientific pipelines.  
Here, we develop a fast quantum algorithmic implementation of an advanced MCMC algorithm.  Given (1) that MCMC is a workhorse algorithm of modern statistical inference and (2) the significant scale of current investment in quantum computing knowledge and infrastructure, other approaches to quantum accelerated MCMC are sure to follow.  We find three particular avenues of future research interesting. 

First, it is clear that the strategies we develop here will provide similar exponential speedups for other advanced MCMC algorithms.  For example, the locally-balanced proposal scheme of \cite{zanella2019informed} generates proposals by selecting among members of a fixed proposal set with probability proportional to the square-root target function.  One may further combine this strategy with other MCMC approaches that encourage fast mixing.  Nonreversible Metropolis-Hastings schemes \citep{turitsyn2011irreversible} maintain momentum between successive MCMC iterations and can lead to orders-of-magnitude faster convergence when sampling from discrete models.  Unfortunately, changing directions in this framework requires a significant number of target evaluations.  Our strategy may confer exponential speedups here as well.  Second, our Ising model case study makes use of local moves, but quantum computers may prove useful for generating proposals far away from the current position in a manner that preserves high probabilities of acceptance.  \cite{layden2023quantum} achieve this but must compute the target probability at the new proposal state from scratch using a conventional computer.  Ideally, one would be able to make global jumps in a manner that uses the quantum device for both proposal and acceptance steps, as we do here.  Finally, the Ising model is a foundational model that one can also use to approximate diverse targets \citep{leng2023quantum}, but unlocking the power of quantum computing for statistics will also require adapting additional discrete models to frameworks like ours.  One powerful possibility is applications that include Bayesian tree-based classifiers and regression models \citep{chipman2012bart,ma2017adaptive}. An open question is whether these methods may also confer exponential speedups when sampling discrete topologies for the trees that underpin these models.

\begin{acks}[Acknowledgments]
This work was supported through National Institutes of Health grants R01 AI153044, R01 AI162611 and K25 AI153816.
PL and MAS acknowledge support from the European Union's Horizon 2020 research and innovation programme (grant agreement no. 725422-ReservoirDOCS) and from the Wellcome Trust through project 206298/Z/17/Z.
PL acknowledges support from the Research Foundation - Flanders (‘Fonds voor Wetenschappelijk Onderzoek - Vlaanderen’, G0D5117N and G051322N) and from the European Union's Horizon 2020 project MOOD (grant agreement no. 874850). AJH acknowledges support from the National Science Foundation (grants DMS 2152774 and DMS 2236854).
\end{acks}



\begin{appendices}

\section{Implementational Correctness}
\label{App. A}.

\begin{figure}[!t]   
    \centering
    \subfigure[Test Graph]{\includegraphics[width=0.3\textwidth]{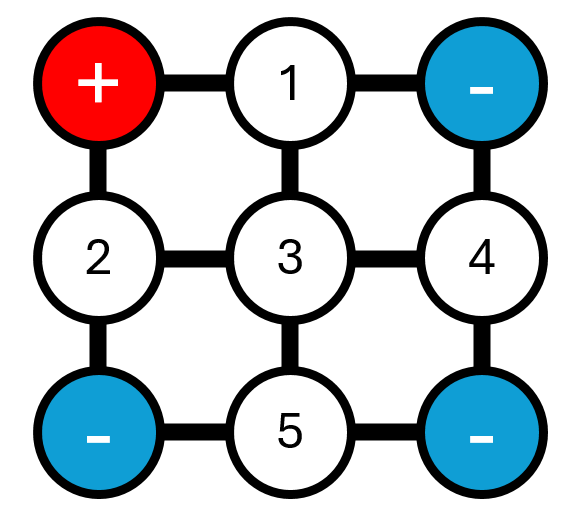}
     }
    \subfigure[Sample Distribution]{\includegraphics[width=0.45\textwidth]{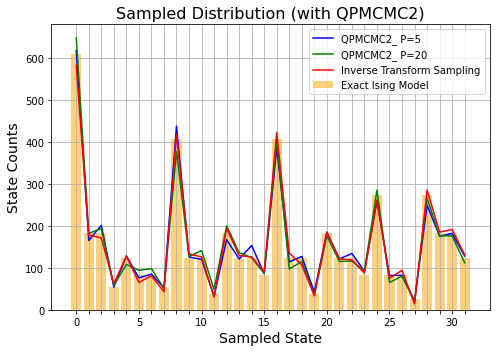}
     }

 \caption{
(a) A $3 \times 3$ square lattice used as our test graph. We designate the four corner vertices as the observed taxa, each fixed with binary values $(+1, -1, -1, -1)$. The remaining five vertices lead to a configuration space of size $|\mathcal{A}| = 2^5$ in our Ising model. 
(b) Distribution of $S = 5000$ samples generated by the \QPP\ algorithm for different numbers of proposals $P$, compared to those generated using inverse transform sampling (ITS) and the exact Boltzmann distribution of the Ising model. 
We show that for different proposal counts $P \in \{5, 20\}$, the samples drawn by \QPP\ closely resemble those generated by ITS and the exact Boltzmann distribution. 
}
 
    \label{fig: small_graph}
\end{figure}

To verify the implementation correctness of our sampling algorithm, \QPP, we test it on a $3 \times 3$ lattice graph Ising model sampling problem and compare the sampled distributions obtained from \QPP\ with those from inverse transform sampling (ITS) and the exact Boltzmann distribution of the Ising model. 
We run the ITS algorithm and \QPP\ to sample the Ising model on a $3 \times 3$ square lattice with four taxa (\cref{fig: small_graph}), using a coupling constant $J = 0.2$. 
The $5000$ samples generated by \QPP, with proposal numbers $P \in \{5, 20\}$, are distributed very closely to the sampled distribution generated by ITS and the exact Ising model distribution. 
The ITS algorithm is implemented using the Python package \texttt{random.choices}.

Next, we fix the number of proposals to $P = 5$ and test different MCMC chain lengths $S$. Let $\text{Algo} \in \{\QPP, \text{ITS}\}$, and we define $\Delta(S, \text{Algo})$ as follows:

\begin{equation}
\Delta(S, {Algo})=\sum_{\alpha \in \mathcal{A}} |\hat{\pi}(S,\alpha,{Algo})-\pi(\alpha )|.
\end{equation}

\begin{figure}[!t]   
    \centering
    \includegraphics[width=0.45\textwidth]{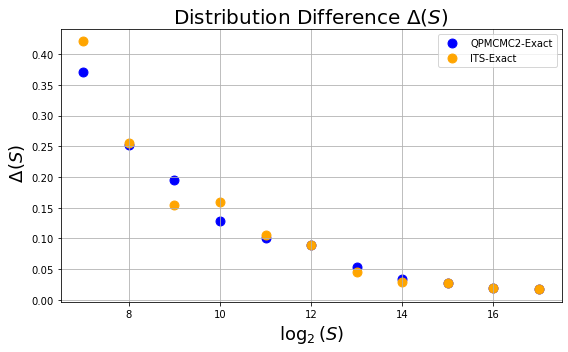}

 \caption{ 
$\Delta(S)$ of \QPP\ (with $P = 5$) and ITS for different chain lengths $S$.  
We observe that $\Delta(S)$ for both \QPP\ (with $P = 5$) and ITS decreases as the chain length $S$ increases, indicating convergence toward the exact Boltzmann distribution.
 }
 
    \label{fig: difference decrease}
\end{figure}

Here, $\mathcal{A}$ is the set of all possible states in our Ising model, with $|\mathcal{A}| = 2^5 = 32$. Given the chain length $S$, $\hat{\pi}(S, \alpha, \text{Algo})$ denotes the sampled distribution of the state $\alpha$ obtained using the sampling method $\text{Algo} \in \{\QPP, \text{ITS}\}$, and $\pi(\alpha)$ represents the exact Boltzmann distribution. The term $\Delta(S, \text{Algo})$ quantifies the difference between the sampled distribution obtained with $\text{Algo}$ and the exact Boltzmann distribution for a given chain length $S$. In \cref{fig: difference decrease}, we observe that both $\Delta(S, \QPP)$ and $\Delta(S, \text{ITS})$ decrease as the chain length $S$ increases, indicating that both sampling algorithms converge toward the exact Boltzmann distribution.

\end{appendices}

 \bibliographystyle{ba}
 \bibliography{references}


\end{document}